# The temporal and spatial scales of density structures released in the slow solar wind during solar activity maximum

E. Sanchez-Diaz[1], A. P. Rouillard[1], J. A. Davies[2], B. Lavraud[1], R. F. Pinto[1], E. Kilpua[3]

[1] Institut de Recherche en Astrophysique et Planétologie, University of Toulouse, CNRS, UPS, CNES, 9 avenue Colonel Roche, BP 44346-31028, Toulouse Cedex 4A, France; eduardo.sanchez-diaz@irap.omp.eu, alexis.rouillard@irap.omp.eu, benoit.lavraud@irap.omp.eu, rui.pinto@irap.omp.eu, illya.plotnikov@irap.omp.eu, vincent.genot@irap.omp.eu

[2] RAL Space, STFC-Rutherford Appleton Laboratory, Harwell Campus, Didcot OX11 0QX, UK; jackie.davies@stfc.ac.uk

[3] Space Physics Department, Department of Physics, P.O. Box 64 FI-00014, University of Helsinki, Helsinki, Finland; emilia.kilpua@helsinki.fi

# Abstract

In a recent study, we took advantage of a highly tilted coronal neutral sheet to show that density structures, extending radially over several solar radii ($R_s$), are released in the forming slow solar wind approximately 4-5 $R_s$ above the solar surface (Sanchez-Diaz et al. 2017). We related the signatures of this formation process to intermittent magnetic reconnection occurring continuously above helmet streamers. We now exploit the heliospheric imagery from the Solar Terrestrial Relation Observatory (STEREO) to map the spatial and temporal distribution of the ejected structures. We demonstrate that streamers experience quasi-periodic bursts of activity with the simultaneous outpouring of small transients over a large range of latitudes in the corona. This cyclic activity leads to the emergence of well-defined and broad structures. Derivation of the trajectories and kinematic properties of the individual small transients that make up these large-scale structures confirms their association with the forming Slow Solar Wind (SSW). We find that these transients are released, on average, every 19.5 hours, simultaneously at all latitudes with a typical radial size of 12 $R_s$. Their spatial distribution, release rate and three-dimensional extent are used to estimate the contribution of this cyclic activity to the mass flux carried outward by the SSW. Our results suggest that, in interplanetary space, the global structure of the heliospheric current sheet is dominated by a succession of blobs and associated flux ropes. We demonstrated this with an example event using STEREO-A in-situ measurements.

# 1. INTRODUCTION

While the fast solar wind is thought to form along the open magnetic field lines rooted at the center of large coronal holes (e.g. Hollweg & Isenberg 2002), the origin of the Slow Solar Wind (SSW) is still the subject of much debate. It has been argued that the SSW forms through a continuous release of transients (e.g. Einaudi et al. 2001; Lapenta & Knoll, 2005; Antiochos et al. 2011). Others (e. g. Wang et al. 2009, 2010) have argued that the SSW consists of two components, one transient and the other continuous, the latter being heated and accelerated via similar processes to those that form the fast solar wind. The properties of the fast and SSW are related to the topology of the magnetic flux tubes channeling the winds (e.g. Wang et al. 2009). In-situ measurements suggest two different sources of SSW, a source at the boundary of coronal holes that produces a SSW with a composition similar to the fast solar wind and a source near the Heliospheric Current Sheet (HCS) that produces a SSW with a low helium abundance and a high abundance of low First Ionization Potential (low-FIP) elements (Kasper et al. 2007, 2012; McGregor 2011; Stakhiv et al. 2015, 2016).

Density enhancements with radial dimensions of several solar radii ($R_s$) emerge continually from coronal streamers and accelerate from less than 100 km s$^{-1}$, close to the inner edge of coronagraph fields of view, up to 300 km s$^{-1}$ at 30 $R_s$, the latter being the typical speed of the SSW in the heliosphere (Sheeley et al. 1997). The frequent release of such structures is considered an important source of SSW. Coronagraph images have revealed that even smaller density structures, with radial extents of the order of 1 $R_s$, also emerge from coronal streamers every 60 to 100 minutes (Viall et al. 2010; Viall & Vourlidas 2015). The advent of high-cadence, in-situ composition measurements has revealed the presence of similar quasi-periodic variations in the abundance of minor ions in the SSW (Kepko et al. 2016). The latter could relate to the smaller density structures observed in coronal images (Viall et al. 2010; Viall & Vourlidas 2015). According to in-situ measurements, 70 to 80% of the SSW is made up of quasi-periodic structures (Viall et al. 2008).

Coronagraph observations suggest that magnetic reconnection is responsible for the release of density enhancements with a periodicity of a few hours (Sanchez-Diaz et al. 2017, Henceforth SD17). We therefore expect these density enhancements to be associated to some magnetic field structure resulting from magnetic reconnection at the solar corona. In a parallel study, the nature of such magnetic structure is being explored with in-situ measurements in order to retrieve further details about the reconnection process at the solar corona. This magnetic structure will be hereafter called "flux rope". We will use the name "blob" for the density enhancement itself, observed in white light images. The whole structure formed by the flux rope plus the blob will be referred to as "small transient".

The direct connection between in-situ measurements and coronal imagery of blobs is possible by tracking the outward motion of density structures in heliospheric imagery, in particular from the Heliospheric Imager (HI; Eyles et al. 2009) instruments on NASA's STEREO mission. This tracking can only be accomplished for those blobs that become entrained by a Stream Interaction Region (called a Corotating Interaction Region, CIR, if it exists for more than one solar rotation; Jian et al. 2006). In this case, a blob's radial expansion is offset by its compression by the fast solar wind (Rouillard et al. 2008). Such compression permits a 3-D mapping of the pre-existing small-scale structuring of the SSW. Rouillard et al. (2010a;b) demonstrated the tracking of blobs all the way to their in-situ detection at 1 AU in combined STEREO observations; the authors showed that these blobs were associated, in in-situ measurements, to the passage of kinks and loops that were likely part of magnetic flux ropes. Similar transient structures had previously been detected in in-situ measurements alone (e.g. Crooker et al. 1996, 2004; Kilpua et al. 2009; Yu et al. 2016). More recently, Plotnikov et al. (2016) mapped blob emission along the ecliptic plane over nearly a complete solar cycle. The authors found that, statistically, blobs emerge from the coronal neutral line.

Provided that the total proton flux of the SSW is known from in-situ measurements at 1 AU (Wang 2010; Sanchez-Diaz et al. 2016), the fraction of the SSW mass flux transported by blobs alone may be obtained by exploiting heliospheric imagery to retrieve the sizes, release rate and spatial distribution of blobs. The tracking of blobs to 1 AU, and their identification in situ, is only possible for certain spacecraft configurations. Moreover, in-situ detection of a transient propagating along a single longitude and latitude does not provide information on the large-scale spatial distribution of blobs. This large-scale picture may, however, be retrieved using remote-sensing instruments. Identification of individual blobs in heliospheric images, and a measure of their release rates, provides information on their spatial distribution and size. Such a mapping of their distribution can provide the context for the interpretation of the in-situ measurements.

This is not easily done when the neutral line - the origin of blobs (Plotnikov et al. 2016) - is oriented in the east-west direction. Because white-light coronal and heliospheric imagers integrate light scattered by all density structures situated along the line of sight, this would include a plethora of blobs distributed in longitude for an imager situated in the ecliptic plane. In this manuscript, we take advantage of the particular coronal magnetic field topology during Carrington Rotation 2137 that was exploited in our previous study (SD17). At that time, the neutral line, along which blobs are continually expelled, exhibited a north-south excursion. In SD17, we exploited the north-south orientation of this neutral line to demonstrate that the formation of blobs is associated with the sunward collapse of loop-like structures. The resulting outward-inward pairs of moving features were related to the effect of magnetic reconnection acting continually to release blobs into the interplanetary medium (SD17). We exploit this period of data further in this paper, to map the size, periodicity of release and spatial distribution of the outflowing blobs. This analysis reveals several new remarkable insights into streamer activity and its contribution to the variability of the SSW.

Blobs have traditionally been observed in running-difference images, constructed by subtracting a previous image from the current image (e.g. Sheeley et al. 1997, 2009). The signature of an outflowing density structure manifests as a white followed by a black feature in gray-scaled versions of such running-difference images. Time-elongation maps (hereafter called J-maps) constructed from running-difference images (e.g. Davies et al. 2009) are a valuable tool in the identification and tracking of blobs (e. g. Rouillard et al. 2008, 2009, 2010a). CIR-entrained blobs can be easily recognized in J-maps constructed with STEREO-A HI images taken prior to the spacecraft's entry into superior conjunction (in 2015), as a series of such blobs emitted from the same source region appears as a characteristic family of converging tracks (Rouillard et al. 2008). In this paper, we combine J-maps with other plot formats that can highlight also the latitudinal extent of blobs, such as time-Position Angle (PA) maps (hereafter PA-maps) and time-latitude maps (hereafter LAT-maps) described later. This combination of maps allows us to observe the space-time distribution and nature of small-scale transients. The J-maps that we will use have been constructed with running-difference data, whereas the PA-maps and LAT-maps have been constructed using background-subtracted data (more specifically, 1-day running background-subtracted data).

The orbital configuration of STEREO and Earth on 2013 June 3 is presented from above of the ecliptic plane depicted in Figure 1a, repeated from Figure 1 of SD17. At this time, the STEREO-A spacecraft was 140° ahead of the Earth. The location of a CIR that passed through the field of view of the HI instruments, and will be presented later, is shown in Figure 1a as a blue spiral. As pointed out by SD17, this CIR was induced by the giant coronal hole evident in Figure 1b, which presents an extreme ultraviolet image of the solar corona taken by the Atmospheric Imaging Assembly (AIA; Lement et al. 2012) instrument on the Solar Dynamics Observatory (SDO; Pesnell et al. 2012). This coronal hole, which passed through the central meridian of the Sun from the viewpoint of Earth on 2013 May 29, is evident as a large dark region near disk center and corresponds to a region of unipolar magnetic field lines connecting the low corona to the interplanetary medium. The presence of this extensive structure forced a strong excursion of the neutral line, which forms at higher altitude, to a north-south orientation as revealed by a Potential Field Source Surface (PFSS) extrapolation of photospheric magnetic fields (Figure 1c). The PFSS extrapolation (Schrijver & DeRosa 2003) is based on evolving surface magnetic maps into which data from the Helioseismic and Magnetic Imager (HMI; Scherrer et al. 2012) onboard SDO are assimilated. Over several days, the CIR, the coronal hole, and the north-south oriented neutral line corotated with the Sun. The CIR was detected by the in-situ instruments on board the STEREO and L1 spacecraft during several Carrington rotations, as will be shown later in this paper and in a future study.

As noted earlier, analysis of the variability at the time scales of blobs along a near-equatorial east–west oriented neutral sheet (i.e., as a function of longitude) is difficult with images taken from the ecliptic plane because white-light features are integrated along the line of sight. By contrast, images of a north–south oriented neutral sheet allow us to study, at a single point in time, the distribution of blobs over an extended surface area of the neutral sheet. The high tilt of the neutral line during this period led to the emission of density structures at all latitudes imaged by HI, as shown in Figure 2. In this paper, we use images from the HI instrument on STEREO-A to derive the 3-D trajectory of density structures emitted by the north-south oriented neutral line, at all latitudes, during its corotation time from the Earth to STEREO-A. We then map the 3-D extent and emission rate of blobs to reveal the large-scale streamer activity. The 3-D mapping of emitted density structures guides us in our search for the in-situ signatures of streamer blobs and, moreover, provides a quantification of their contribution to the SSW mass flux. Finally we discuss the implication of our findings in furthering understanding of the formation and evolution of the SSW and, more generally, the large-scale coronal magnetic field.

## 2. OBSERVATIONS

### 2.1. Trajectories of Blobs Derived from J-maps

The left-hand column of Figure 3 displays three J-maps constructed using HI-1/2A running-difference images at three different PAs during the passage of the CIR of interest, from 2013 May 28 to June 06. No CMEs are observed during this period and the variability of these maps is dominated by tracks that are the signature of the outward-moving blobs that are entrained by the CIR.

The central column repeats the J-maps in the first column but with, overlaid as blue curves, the result of a trajectory analysis of the manually-extracted time-elongation profile of each blob using the technique described by Rouillard et al. (2010a). This technique assumes radial propagation at a constant speed in a constant direction. For each blob, which is assumed to be a point source, the analysis provides estimates of:

> (1) Its direction of propagation within the plane corresponding to the selected PA, defined by the angular separation ($\beta$) between the propagation direction of the blob and the Sun-observer line, the observer in this case being STEREO-A (see Figure A1 in the Appendix).
> (2) Its radial velocity in this direction (v), and
> (3) The time at which it passed through an arbitrary height of 20 $R_s$ ($t_0$).

These three parameters are tabulated on the website of the EU FP7 HELCATS project (https://www.helcats-fp7.eu) for each blob observed during the passage of the CIR through STEREO-A/HI J-maps, for PAs from 50° to 130° separated by 2°. The median speed of all blobs during this period was found to be 390 km s$^{-1}$. This confirms that they are propagating in the SSW. In each HI J-map, we can see that both the real tracks (left-hand column) and the fitted tracks (central column) tend to converge, a characteristic signature of blob emissions from a single corotating source region on the Sun as viewed from the vantage point of STEREO-A prior to superior conjunction (Rouillard et al. 2008). The properties of those blobs propagating in the ecliptic plane that constitute the ecliptic sub-structure of the CIR were derived by Plotnikov et al. (2016) and are listed in the CIR catalogue (so-called CIRCAT) produced by these authors for the FP7 HELCATS project.

We can compare this converging pattern with the one that would result from the emission of blobs from a corotating source region that is defined by the position of the neutral line. For each track identified as a blob in the trajectory analysis, we recalculate the $\beta$ angle of the blob corresponding to the position of the neutral line at the time of release of that blob at 1 $R_s$ (SD17). A common velocity of 390 km s$^{-1}$, the median speed of all fitted blobs, is assumed, from which that release time is calculated. The resulting pattern is overplotted in the panels in the right-hand column of Figure 3. The match between the observed and modeled pattern of tracks confirms that the blobs are released from near the neutral line.

## 2.2. Estimation of the Radial Size and Rate of Release of Blobs

To study the distribution of blob emission with latitude, we have built a LAT-map that displays, as a function of time, the white light that is Thomson scattered by electrons that are estimated to be located at a heliocentric distance of 30 $R_s$. In contrast with the running-difference images, the 1-day background-subtracted images that are used to build the LAT-map shown in Figure 4a are much less sensitive to the motion of density structures. While running-difference images are particularly useful at highlighting moving density features, background-subtracted images are much better suited to observe the real extent and spatial distribution of density structures. To derive the heliocentric distance, we assume that, during this otherwise very quiet time interval of solar activity, the scattered light comes exclusively from the CIR-entrained blobs. We further assume, as above, that each blob moves along a fixed heliocentric latitude and longitude defined by the location of the neutral line at the release time of the blob at 1 $R_s$. All blobs are attributed a speed of 390 km s$^{-1}$ in this direction (the median speed of the trajectories of all blobs during this period). Due to solar rotation, a family of such blobs will form a spiral in the interplanetary medium rooted at the neutral line near the Sun; hence, in essence, the LAT-map is constructed from those pixels in the heliospheric image, the lines of sight of which intersect the interplanetary spiral defined by a speed of 390 km s$^{-1}$ at a height of 30 $R_s$ above the Sun. The coordinate transformation from angular units of the instrument (PA, elongation) to the heliographic equatorial coordinates (HEEQ) of the blob (height, latitude, longitude) are described in detail in Appendix 1. The LAT-map displays the brightness at a fixed height of 30 $R_s$, as a function of time (X-axis) and latitude (Y-axis).

Figure 4a reveals a remarkable sequence of five bright vertical bands, nearly equally spaced in time, separated by dark bands. These bright bands are indicated in the Figure with five blue rectangles. They are labeled with roman numbers from I to V, sorted by date and time. The bright bands correspond to density structures emitted simultaneously at all latitudes. The dark bands, which also extend over all latitudes, mark periods of low-density emissions of similar duration. Figure 4b presents a time series obtained by averaging brightness from all latitudes. The alternating periods of high and low-density emissions are clearly reflected in that time series, exhibiting a clear periodicity; each peak in brightness in Figure 4b coincides with a bright vertical band in Figure 4a. Figure 4 illustrates, for the first time, a clear tendency for the streamer belt, and its embedded neutral line, to produce an emission over an extended region of the streamer belt. The physical mechanisms for this remarkable property will be discussed later; we first extract more information on this emission process.

To that end, we performed a Fourier spectral analysis of the brightness time series in constant latitude bands (horizontal slits) of the LAT-map. Figure 5 shows the magnitude (top panel), and phase (bottom panel) of each of these Fourier transforms as a function of latitude (Y-axis) and frequency (X-axis). We fitted each spectrum (the magnitude of the Fourier transform) to a red noise, i. e., a noise whose spectral density decreases with decreasing frequency according to the function $S=S_0 \cdot f^\alpha$, where $S_0$ and $\alpha$ are constants to be fitted and S is the spectral density of the noise. The noise was subtracted from each spectrum; such noise-subtracted spectra are plotted in Figure 5a. Nearly all latitudes between 20° south and 30° north exhibit a peak in spectral magnitude with a periodicity of 19.5 hours. The phase of the Fourier transform (Figure 5b) that corresponds to this periodicity is nearly constant with latitude; the density structures reach an altitude of 30 $R_s$ simultaneously, at all latitudes, every 19.5 hours. Hence, if all of these density structures move at roughly the same speed, this demonstrates that the release of transients occurs simultaneously at all latitudes, every 19.5 hours.

Given that all of the 1D brightness-time series exhibit the same peak in periodicity at 19.5 hours, with the same angular phase, we can also investigate the periodicity of the time series obtained by averaging all latitudes (shown in Figure 4b). Figure 6 shows the Fourier transform of the latitude-averaged brightness-time series at 30 $R_s$ (panel a), as well as the Fourier transform of analogous time series constructed at altitudes of 40 $R_s$ (panel b) and 50 $R_s$ (panel c) above the solar surface. The magnitude of the Fourier transform is plotted as a black solid line; the red dashed line corresponds to a 99% confidence level, according to the chi-square test, above the red noise spectrum $S=S_0 \cdot f^\alpha$ that best fits the frequency spectrum of the brightness-time series. The angular phase associated with the spectral peak corresponding to the bright bands is shown as a blue asterisk.

The spectral peak at 19.5 hours, which was visible in Figure 5a, is readily apparent in the Fourier transform of the latitude-averaged brightness-time series at 30 $R_s$ (Figure 6a). A spectral peak at the same periodicity is also present at 40 and 50 $R_s$ although less prominent compared to the noise than that at 30$R_s$. The phase associated with the spectral peak provides information on the time difference, $\Delta t$, between the first brightness peak and the chosen time origin. This time difference can be calculated as:

$$\Delta t = -\frac{\delta}{360°} T$$

where $\delta$ is the angular phase of the Fourier transform and T, the periodicity of the spectral peak. The decrease in the phase with altitude demonstrates that there is a time shift between the different altitudes, which corresponds to the time taken for the density structures, observed as vertical bands in Figure 4a, to travel between them. The phases associated with the spectral peak at 19.5 hours are 100°, -35° and -100° at 30, 40 and 50 $R_s$, respectively. This translates to -5.4h, 1.7h and 5.8h time shifts. Taking into account the separation of 10 $R_s$ between subsequent brightness time-series, our analysis reveals that the average speed of these structures was 250 km s$^{-1}$ between 30 and 40 $R_s$ and 450 km s$^{-1}$ between 40 and 50 $R_s$ revealing that the density structures are accelerating between 30 and 50 $R_s$. The terminal speed of 450 km s$^{-1}$ is very close to the speed of the SSW measured in-situ a few days later by STEREO-A. Therefore, we infer that the bright vertical bands of Figure 4a are associated with structures that are advected in the SSW. Furthermore, the median speed of the blobs derived from the J-maps (390 km s$^{-1}$) lies in between the two speeds calculated via this analysis. This suggests that the bright vertical bands in the LAT-map shown in Figure 4a are indeed the signatures of the same blobs that are observed as the traces in the J-maps.

We also construct PA-maps, like the LAT-maps, using 1-day background-subtracted images, but remaining in a spacecraft-centered angular coordinate system (helio-projective, HPR, radial coordinates: i.e. PA and elongation). The PA-map is built by extracting information at a fixed elongation of 5° (rather than at a fixed height of 30 $R_s$ as done in the LAT-map shown in Figure 4a). Figure 7 compares the time at which the blobs passed through 5° elongation in the J-maps with the position of the bright vertical bands in a PA-map at 5° elongation. Both PA-maps and J-maps exploit the same HPR coordinate system. Each panel in Figure 7 displays, underneath, a J-map constructed at a given PA and, above, a portion of the PA-map centered at the same PA (specifically 80° and 90° for Figures 7a and 7b, respectively). The trajectories of blobs, calculated from their manually-extracted time-elongation profiles, are shown as blue solid lines in the J-maps and their times of passage through 5° elongation are shown with blue dots in the PA-map.

Figure 7 reveals that the tracks in J-maps that are typically attributed to outflowing density structures (or blobs) entrained by a CIR tend to coincide with the boundaries of bright vertical bands observed in the PA-maps. Only a few blobs map to the central portions of the vertical bands. Therefore, we suggest that the tracks considered in the J-maps, which are built from running-difference images, map the evolution of the outflowing bright bands containing density enhancements. We conclude (1) that the brightness distribution observed in background-subtracted images maps the actual density distribution in the image plane, (2) that the vertical bands seen in PA-maps track outflowing packets of blobs, and (3) that the blobs in the running-difference J-maps reveal the individual density variations contained within that outflowing structure.

SD17 determined how the tracks observed in the J-map of Figure 7a originated in the corona by analysing STEREO COR2 images. SD17's observations show that the tracks appear during transient events that involve the release of material in the corona and the associated sunward collapse of material, in a process called 'raining inflows' (e. g. Sheeley & Wang 2001, 2014). The observations by SD17 were indicative of the process of magnetic reconnection acting to release transient structures in the SSW just above helmet streamers. Such a detachment was especially clear in Figure 4 of SD17 for the track marked by the first blue dot in Figure 7a. This track lies at the leading edge of the first bright vertical band (labeled band I in Figure 4), i. e., at the leading edge of a blob.

Here, we also investigate the link between features observed in the HI-1 images themselves with those observed in our different types of maps. Figure 8a shows the PA-map at 5° elongation extending from 2013 May 31 to June 3. The green and red lines mark the passage, through 5° elongation, of two of the tracks observed in the J-maps (Figure 7) that, as already concluded, lie at the edges of the bright vertical bands in the PA-map. In this case, they lie at the edge of the band IV. The locations of these blobs at 10:49 UT on 2013 June 02 are shown in a running-difference (Figure 8b) and a background-subtracted (Figure 8c) image as correspondingly-colored green and red dots. The area between the red and green dots corresponds to an extended region of enhanced brightness in Figure 8c that seemingly consists of distinct structures in Figure 8b. Running-difference images, useful for highlighting brightness variations associated with moving features, cannot provide unambiguous information on the nature and extent of density emissions released in the SSW. Figure 8c confirms that each bright band is composed of multiple narrow sub-structures that extend mainly along a solar radial, a feature that is already evident from the LAT-maps. Later, we will argue that each of these sub-structures corresponds to material carried by a magnetic transient released from the streamer, probably by magnetic reconnection (SD17).

Previous studies have found periodicities in blob emission ranging from 2-3 blobs per day (Rouillard et al. 2010a) to 4-6 blobs per day (Wang et al. 1998, Song et al. 2009). These periodicities account for the number of brightness enhancements observed in running-difference images. We have shown with Figures 7 and 8 that a single blob may typically be associated with two or more brightness variations in running-difference images, corresponding to at least the leading and trailing edge. Therefore, the frequency of release of blobs may have been overestimated in previous studies. Moreover, the apparent size of a blob is larger in a background-subtracted than in a running-difference image.

In order to better determine the radial extent of blobs, we calculate the wavelet transform of the latitude-averaged brightness-time series at 30 $R_s$. We remind the reader that this time series is calculated as the average over all latitudes in the LAT-map shown in Figure 1a (i. e. for all the data available between -60° and 60° latitude), which was constructed with background-subtracted images. Hence we neglect any latitude variation and assume that the duration of the band is exactly the same at all latitudes. As a mother wavelet, we use the Gaussian wavelet of order 4. Figure 9a shows the latitude-averaged brightness-time series itself (replicating Figure 4b) and Figure 9b shows its wavelet transform as a function of time (X-axis) and wavelet scale (Y-axis). The wavelet transform shows periodic peaks at scales between 5h and 15h with a periodicity that is similar to the periodicity found for the vertical bright bands in Figures 4a.

In order to find the wavelet scale that best fits the periodicity of the blobs, we calculate the frequency spectrum at each scale of the wavelet transform. These spectra are shown in Figure 9c. The color of each line in Figure 9c corresponds to the wavelet scale in a gray scale from 40 min (white) to 26h (black). The frequency spectrum with the most prominent peak, at a period of 19.5 hours, is the one corresponding to a scale of 9h. The spectrum of the wavelet transform at scale 9h is shown as a thicker blue solid line in this panel. 9h is, therefore, the time that it typically takes for each blob to pass over a fixed height of 30 $R_s$. The spatial size of blobs in the radial direction can be calculated as this time multiplied by their speed. Using a speed of 250 km s$^{-1}$, which is the average speed of blobs between 30 and 40 $R_s$ (Figure 6), their size in the radial direction at 30 $R_s$ is estimated as being 12 $R_s$.

## 2.3. Latitudinal Distribution of Blobs

Each band of enhanced brightness actually exhibits considerable variability with latitude (Figure 8c). In order to find the latitudinal distribution of blobs, we repeat our spectral analysis on brightness-PA profiles taken at fixed times. Figures 10 and 11 show the results of such analysis for all such brightness-PA series contained in two of the bands of enhanced brightness.

In Figure 10, we focus on the 6 hours time interval extending from 2013 May 30 18:00 UT to May 31 00:00 UT, which corresponds to the passage of the first vertical band (band I) in Figure 4a. Figure 10a shows the brightness versus PA profile from 50° to 130° PA from each image within the time period of band I, as thin solid lines of different colors. The thick black solid line presents the average of all brightness-PA profiles in band I. Figure 10b presents the wavelet transform of this averaged brightness-PA profile. At longer wavelet scales (10°-20°), four quasi-periodic increases in wavelet power are observed, at a rate of one every 30°-40° of PA. They correspond to the four bright structures of larger spatial scale in Figure 10a. The shorter wavelet scales (5°-10°) also show quasi-periodic peaks in wavelet density, which correspond to the sub-structure contained within each of these large-scale structures. This reveals that, contrary to the time distribution that is dominated by a single periodicity of 19.5 hours associated with the scale of 9h (see Figures 5-6), the latitudinal distribution of small blobs involves several scales and periodicities. Figure 10c presents the frequency spectrum at each scale of the wavelet transform shown in Figure 10b. Note that we have again subtracted a red noise from the spectrum, with the same method used to produce Figure 5. The color of each line in Figure 10c corresponds to the wavelet scale, as a gray scale from 1° (white) to

20° (black). The wavelet scales between 15° and 20° are responsible for the spectral peak extending from 29° to 44° PA while the spectral peak with a periodicity of around 18° in PA is generated by the scale at 8°. Note that 29° and 44° are the second and third points on the X-axis of each spectrum, as each spectrum is 88° long. Therefore the spectral resolution does not allow us to discern any intermediate frequency between 29° and 44°.

Figure 11 is analogous to Figure 10 but for the time interval extending from 2013 June 2 00:00 UT to 08:00 UT, which corresponds to the passage of the fourth vertical band in Figure 4a (band IV). Figure 11a shows the brightness versus PA profile from each image within the time period of band IV, and their average (solid line). The wavelet transform of the averaged brightness-PA profile of band IV, displayed in Figure 10b, highlights again the multi-scale nature of the latitudinal distribution, with four large-scale brightness enhancements and additional smaller-scale peaks. The spectrum at each scale of the wavelet transform is shown in Figure 11c. In this case, the wavelet scale of 13° forms a prominent spectral peak at 29° PA. The wavelet scales of 7° to 8° generate an additional, but less prominent, peak at 15° periodicity in PA. Even though the multi-scale nature of the latitudinal distribution is similar for both bands I and IV, the scales involved and, in particular, their relative importance are not the same.

Table 1 shows the scales and periodicities corresponding to the latitudinal distribution of each band. In general, a small wavelet scale of 7° to 12° is associated with a periodicity of 15° to 20° in PA and a large scale, of 15° to 20°, is associated with a spectral peak with a periodicity of 29° to 44°. We directly examined the HI1-A background-subtracted images to identify peaks associated with wavelet scales of 7° to 12° in the brightness-PA profiles. We find that such peaks (associated with a periodicity of 15° to 20°) systematically correspond to the small-scale brightness enhancements that are usually associated with blobs (e. g. in Figure 8). We can, therefore, conclude that, at a height of 30 $R_s$ above the Sun, blobs are spaced every 15° to 20° in PA and have a PA extent of 7° to 12°.

## 2.4. Spatial and Temporal Distributions of Blobs

The spatial and temporal scales of density inhomogeneities released from a highly tilted neutral line have been determined with spectral and wavelet analysis of heliospheric images. The results of this analysis are summarized in the schematics presented in Figure 12, which illustrate the spatial distribution of small transients in the heliosphere at any given time. Figure 12a shows a latitudinal cut through the density enhancement, blob, associated with a single, typical, small transient. It can be represented as an ellipse with major and minor axes of dimension 12 $R_s$ and 5 $R_s$, respectively (see section 2.2).

Figure 12b illustrates, for a given time and in a plane of constant heliographic longitude, the latitudinal and radial distribution of blobs (grey areas), defined as high density regions, released by a highly tilted neutral line. The common origin of outflowing blobs and raining inflows (SD17) suggests that magnetic reconnection occurs at the origin of such outward-moving density enhancements (SD17). For this reason, we expect the high-density latitudinally-extended structures, observed as bright vertical bands in the LAT-map (e.g. Figure 4), to be associated with weak magnetic fields followed by flux ropes that fill the intervening dark bands in the LAT-map (e. g. Figure 4). Depending on how reconnection takes place at the origin of blobs, the associated magnetic structures might be magnetically connected to both hemispheres of the Sun, be connected only to one point in the solar corona, or constitute a totally disconnected plasmoid. We will use in-situ measurements, in next subsection and in a future study, to investigate the magnetic connectivity of these magnetic structures in more detail. Regarding their latitudinal distribution, our results suggest that blobs are typically separated by 15° from one another and have a latitudinal extent of 10° at 30 $R_s$, which corresponds to 5 $R_s$ (see section 2.3). The apparent coalescence, in images, of these structures would generate larger-scale brightness enhancements spaced every 30° in latitude, with a latitudinal extent of some 15° to 20°. Viall & Vourlidas (2015) showed that small-scale blobs are made up of even smaller structures that only extend over about 1$R_s$; it is therefore likely that even the brightness peaks that we identified using our spectral analysis with periodicities of 15° PA and wavelet scale of 10°, are made of the coalescence of even smaller-scale blobs.

Finally, Figure 12c shows, for a given time, the spatial distribution of small-scale transients along the Parker spiral in a plane of constant heliographic latitude. The stream interface would consist of a succession of density structures (blobs), likely with low magnetic field (gray areas in Figure 12c) separated by magnetic structures. The brightness bands are slightly wider than the intervening darker bands. The blobs observed in running-difference images are thus unevenly spaced in time, with their periodicity alternating between 9 hours (from the leading to the trailing edge of the same density structure) and 10.5 hours (from the trailing edge of a density structure to the leading edge of the next one). During the period of study presented in this paper, four brightness enhancements are observed in 29 hours in running-difference images, i.e. 3.3 per day, corresponding to the leading and trailing edges of two blobs in background-subtracted images. Note that this number is consistent with the periodicities found by Wang et al. (1998) and Song et al. (2009) in running-difference images.

## 2.5. In-situ Measurements

The black and red arrows in Figure 12b represent the possible trajectories of an ecliptic-orbiting spacecraft crossing a north-south oriented neutral sheet. We can identify five potential scenarios:

> **Scenario 1 (blob + flux rope):** The spacecraft crosses the sector boundary by first intersecting a flux rope and then a blob. A spacecraft intersecting the edge of the flux rope in this way would measure the poloidal magnetic field component of the rope only. According to Figure 12, if the central axis of the flux rope is vertical, the spacecraft would measure a rotation of the magnetic field primarily contained in a plane of constant latitude. The intersection through a blob would be manifest in in-situ measurements as a high-density region with highly variable density and magnetic fields. The magnetic fields inside the blob could be lower than in the flux rope as a result of magnetic field annihilation due to reconnection occurring in the corona.
>
> **Scenario 2 (HCS):** The spacecraft crosses a sector boundary characterized by a single, narrow current sheet (the HCS). This 'clean' HCS occurs when the neutral line has had time to reform and thin out prior to the release of a new flux rope from the tip of the helmet streamer. The expected signature is that of a (nearly) concomitant reversal of the magnetic field polarity and true sector boundary marked by a clear reversal in the pitch angle distribution of suprathermal electrons. The region surrounding the HCS could be a narrow region of elevated densities located at the polarity inversion, typically referred to as the HPS (Heliospheric Plasma Sheet).
>
> **Scenario 3 (blob):** The spacecraft crosses the center of the blob. The expected signature is a broad high-density region lasting up to 9 or 10 hours, correlated with weaker but complex magnetic fields.

**Scenario 4 (flux rope + blob):** The reverse trajectory to scenario 1. The spacecraft crosses the edge of the magnetic flux rope and then the edge of a blob. In this scenario, the in-situ measurements would exhibit first those of a flux rope crossed near its edge and then the signature of a blob.

**Scenario 5 (flux rope):** The spacecraft crosses the magnetic flux rope replacing the HCS, with no signature of a blob. Note that the whole streamer, at the tip of which the blobs and flux ropes are forming, is itself a dense structure but the image processing subtracts the background brightness of the more stable streamer in order to highlight the blobs. Therefore, even flux ropes could exhibit relatively high densities but, according to the results of section 2.2, necessarily lower than those measured in blobs.

We note that the longitudinal separation between the emission of two consecutive small transients is typically 12°, which corresponds to 19.5 hours of solar rotation. This time is much longer than the typical plasma sheet crossing time for a spacecraft that crosses the neutral sheet in a perpendicular direction. Geometry makes it extremely unlikely for a spacecraft orbiting at constant heliographic latitude to observe two blobs when it crosses a highly tilted neutral sheet.

These results set the context for analyzing the in-situ measurements corresponding to the crossing of a north-south oriented neutral sheet and the search for in-situ signatures of blobs. During the period of study, the CIR that entrained the density structures observed in Figures 3 to 11 was located between STEREO-A and Earth. This CIR corotated to the location of STEREO-A over the next few days. A large CME, directed towards STEREO-A, was observed by the SOHO and STEREO coronagraphs, which impacted STEREO-A at the same time as the CIR passed over the spacecraft. This prevented this CIR from being clearly detected in-situ by STEREO-A during Carrington rotation 2137. Unfortunately, this was also the case for the subsequent passage of the same CIR over STEREO-B and the L1 point (during Carrington rotation 2138).

Fortunately, however, no CME impacted STEREO-A during the passage of the CIR over that spacecraft during Carrington rotation 2138. The in-situ measurements corresponding to the passage of this CIR over STEREO-A are shown in Figure 13, which extends from 2013 July 8 UT to 11 UT. The passage of a small flux rope associated to a small transient is delimited by the first and second vertical black dashed lines. The passage of the plasma sheet occurs directly thereafter, between the second and third vertical black dashed lines. The plasma sheet contains the neutral sheet, which is characterized by a 180° reversal in the suprathermal electron pitch angle distribution collocated with a reversal in the radial component of the magnetic field (Crooker et al 2004; Owens et al. 2013).

Embedded in the plasma sheet, there is a density enhancement of one to two hours duration that coincides in time with a period of low magnetic field magnitude. It is preceded by the flux rope, with enhanced magnetic field magntitude. The duration of this flux rope (8 hours) is similar to the duration of the flux ropes measured by Rouillard et al. (2010b). This whole signature is consistent with a crossing of the type described in Scenario 4 above. The probe first encounters the edge of a flux rope, since the magnetic field does not show a full rotation as would be expected for a trajectory through the center of a flux rope, and then a density enhancement. A depletion of the suprathermal strahl is partially coincident with the density enhancement. This suggests that the blob itself is totally disconnected from the Sun (McComas et al. 1989, Gosling et al. 2005 Pagel et al. 2005, Crooker & Pagel 2008). Note that complete disconnection is evidenced by the fact that the strahl disappears in both the parallel and anti-parallel directions. The flux rope conserves the polarity of the preceding ambient solar wind; it is, therefore, assumed that it is still attached to one hemisphere of the Sun by one leg, but that the other leg has probably reconnected.

# 3. DISCUSSION

SD17 linked the continual outflow of mesoscale density structures to the sunward motion of density structures known as raining inflows (Sheeley & Wang 2002). This is a strong indication that magnetic reconnection, occurring near 5 to 6 $R_s$, initiates the release of density structures in the SSW. This height of reconnection, estimated in SD17 near solar maximum, may vary for different streamer structures between solar minimum and solar maximum. The combination of the classic running-difference J-maps, as used by authors such as SD17, and new map formats, based on background-subtracted images, reveals that the tracks previously observed in running-difference images, and interpreted as individual blobs, are the boundaries of more extensive high density structures. These structures correspond to the simultaneous release of blobs from extended portions of helmet streamers. We have revealed that the streamer can be activated synchronously over a broad region that is 7 to 8 times broader in latitude than the typical extent of blobs. When the streamer is highly tilted, the profusion of blobs suddenly released in the SSW creates a broad latitudinally-extended structure. The two clearest inflow-outflow pairs were marked with white arrows in Figure 3 of SD17; they are associated with the leading edge of this dense structure rather than its trailing edge.

## 3.1. Spatial and Temporal Scales for the Release of Blobs into the Slow Solar Wind

We have found that, during this solar maximum period, the periodicity of release of small transients is 19.5 hours, and that they typically take 9 hours to cross a particular line of sight. Running-difference images do not provide an easy way to isolate the extended passage of a blob because they highlight only leading and trailing edges. The manifestation of blobs in running difference J-maps is as a sequence of tracks typically separated by 9h (between the leading and trailing edges of a blob) and 10h30 (between the trailing edge of a blob and the leading edges of the subsequent blob). The separations between the consecutive brightness enhancements in running-difference images analyzed by Rouillard et al (2010a), and listed in Table 1 of that paper, are 19h, 11h, 15h, 10h, 12h for CIR-D and 9h, 11h, 6h, 10h for the first five blobs of CIR-E. These values are derived from the time at which each blob/signature passed through an elongation of 5°. Even though the periods vary from one CIR to another, the same pattern of alternating long and short periodicities remains. Note that there is more uncertainty in the periodicity for the last feature, in particular, associated with each CIR. This is because the height above the Sun depends critically on the β angle for a fixed elongation. In this case, β decreases with time for consecutive features within each CIR viewed by HI on STEREO-A. The height at 5° elongation dramatically increases with decreasing β when β<45°, which is the case for the last feature of each sequence (see Equation 2 of Rouillard et al. 2010a).

Note that, if blobs were homogeneous density structures, the white-light signatures of their trailing edge would be different from their leading edge's signature in running-difference images. In such case, the signature of the leading edge of a blob in a running-difference J-map would be a bright track, indicating a density increase from the time prior to the blob passage to the beginning of the blob passage. The trailing edge, on the other hand, would manifest as a dark track, indicating the decrease in density at the moment when the blob leaves the pixel. Most tracks in the J-maps were, however, bright traces. They therefore track increases rather than decreases in density. Bright tracks in the J-maps are therefore associated to some substructures of the blob, mostly close to their edges, rather than associated to the edges themselves. An on-going survey of the in-situ measurements of blobs is meant to analyze such structuring of the density inside blobs.

2D MHD simulations have reproduced the release of transients from an unstable streamer with a periodicity of several hours (Endeve et al. 2003, 2004, Allred & McNeice 2015), and a typical size of 1 to 2 $R_s$ directly after their formation (Endeve et al. 2003, 2004). These simulations suggest magnetic disconnection of blobs from the Sun. The LAT-map presented in Figure 4 reveals that such release of small transients occurs simultaneously at all latitudes (Figures 4-6). Understanding this simultaneity, all along the neutral line requires 3D simulations of the coronal magnetic field including the entire neutral line.

Our present study revealed, thanks to the analysis of background-subtracted images of the heliosphere, the characteristic spatial and temporal scales for the release of blobs during solar maximum. This opens the question of what physical mechanisms could be regulating such particular time and spatial scales, i. e., why are blobs released every 19.5 hours and not other periodicity and why is their size on the plane of sky 12x5 $R_s$ and not any other size. A full detailed answer to these questions is beyond the scope of this paper, but we can find in the scientific literature some candidate mechanisms that could give the observed shape to the blobs or regulate the periodicity of release.

Roudier et al. (2014) studied the horizontal movements of supergranules in the chromosphere and found they have an average life-time of around one day, which is very similar to the periodicity of release of blobs found here (19.5 hours). The movements of plasma on the solar surface determine the motion of the footpoints of coronal loops. It is possible thus that the plasma movements at the supergranular scale could periodically bring the coronal loops closer to each other below the coronal streamer. As a consequence, these coronal loops could reconnect, in a similar picture as the one proposed by Gosling et al. (1995) for the onset of CMEs, producing at the same time the release of blobs and the merging and restructuring of supergranules. Combined chromospheric observations of the supergranular motions and coronal observations of the coronal loops could be useful in the future to explore this hypothesis.

Regarding the shape of blobs, if we assume that blobs are created through magnetic reconnection, as suggested in SD17, then we might expect a strong asymmetry in the shape of blobs right after their formation, 5 to 6 $R_s$ above the center of the Sun (SD17). The aspect ratio of a plasmoid forming as a consequence of magnetic reconnection is equal to the dimensionless reconnection rate, which is typically 0.1 (Cassak & Shay 2007 and reference therein). However, the shape of blobs observed at a heliocentric distance of 30 $R_s$ is 12x5 $R_s$ in the plane of sky, i. e., their aspect ratio is 0.4. Therefore, they must have either been compressed in the radial direction during their propagation in the heliosphere, while keeping approximately their original latitudinal extent. This expansion of blobs into a "pancake" shape was already observed by Rouillard et al. (2008) using running-difference images and also occurs in some CMEs (e. g. Crooker & Horbury 2006). As we shall see in a following study in-situ measurements and those by Rouillard et al. (2010b) suggest that blobs are surrounded by complicated magnetic flux ropes, often with strong and rotating magnetic fields. The magnetic pressure imposed by such flux rope structures could be responsible for the compression of blobs along the radial direction.

In addition, the images show that the brightness enhancement that constitutes a blob may consist of smaller structures. Our estimate of a radial extent of 12 Rs is therefore for the total brightness enhancement. Blobs could result from the development of multiple reconnection sites producing several plasmoids. In a future study, we will show that each substructure is separated by small-scale twisted magnetic field. This smaller-scale variability exhibits fluctuating values of the plasma beta that are likely to affect the global expansion rate of the blob.

### 3.2. Detection of Blobs In-situ

During the period of study, idealized blobs can be projected onto the plane of sky as ellipses with major and minor axes of 12 $R_s$ and 5 $R_s$, respectively. It is not possible to directly resolve the size of small blobs along the line of sight, which in this case corresponds to the azimuthal direction, with images taken from a single vantage point. A precise calculation of the size of blobs along the direction normal to a highly-tilted neutral line, would require the observation of blobs in the heliosphere from multiple viewpoints simultaneously, which is not feasible for Carrington rotation 2137. The angular separation between STEREO-A and B during this time period is too large to observe the same blob simultaneously from both spacecraft (see Figure 1a). In the case presented by Rouillard et al. (2008), which corresponded to blobs released by a more horizontally-aligned neutral line, the latitudinal direction is more closely perpendicular to the neutral line. One blob that is clearly observed by Rouillard et al. (2008; see their Figure 2) has a PA size of 14° at 15° elongation. Its PA size increases with increasing elongation angle, at a rate of approximately 2° per 5° elongation out to 15° elongation (probably due to the compression in front of the CIR, see Section 3.1) and remains the same thereafter (out to 24°). Therefore, its latitudinal size at 5° elongation (approximately 30 $R_s$) would be 10°, which is equivalent to 5 $R_s$.

Even without an accurate estimate of the azimuthal size of blobs, i.e., their size along the direction of the line of sight, our analysis of their variability along the neutral line, based on remote-sensing data, sets the context for the analysis of the in-situ measurements that are presented in Figure 13. In an ongoing study, we are analyzing the in-situ data of similar HCS crossings in the context of what we have learned from this study. One constraint is imposed by the geometry. Since the periodicity of release of blobs is around 19.5 hours, which is much longer than the time that a spacecraft at 1 AU would spend crossing a north-south oriented HCS, such a spacecraft would never detect two blobs during a single crossing of such a HCS. For such a crossing, it is possible for the spacecraft to (1) detect first the edge of a density enhancement (blob) and then the edge of a magnetic structure; (2) cross a classical clean HCS; (3) cross only the density enhancement; (4) cross first the edge of the magnetic structure and then the density enhancement; or (5) cross only the flux rope, missing entirely the density enhancement. Scenario 4 appears to be the case for STEREO-A's crossing of the neutral sheet studied here, from the in-situ measurements presented in Figure 13. These in-situ measurements of a small transient embedded in the plasma sheet raise the question of whether the blobs are embedded in a larger scale structure of high density that constitutes the plasma sheet or whether the plasma sheet measured in-situ is made of a blob or a succession of blobs. We will address this question also in this upcoming study with the in-situ measurements of several sector boundary crossings. Passing through two blobs would be possible for a spacecraft that remains close to the neutral line for longer than 19.5 hours. This is more likely to be the case for a relatively flat east-west directed neutral line. The second clue to aid interpretation of in-situ measurements comes from the size of blobs derived from wavelet analysis.

### 3.3. Contribution of Blobs to the Total Mass Flux of the Slow Solar Wind

The typical volume of a blob at 30 $R_s$ can be calculated as V=(4 $\pi$ /3)•6•2.5•2.5 $R_s^3$ = 157 $R_s^3$. If we assume that such blobs have typical SSW densities of 10 $cm^{-3}$ at 1 AU, which will correspond – with an $R^{-2}$ expansion – to 515 $cm^{-3}$ at 30 $R_s$, each blob would contain 2.7 $10^{46}$ protons. If we consider that one blob is released every 15° along the whole neutral line, then 24 such blobs are released every 19.5 hours, each containing 2.7 $10^{46}$ protons. This means that 6.5 $10^{47}$ protons, in the form of blobs, cross a spherical surface of radius 30 $R_s$, centered on the Sun, every 19.5 hours, i. e., 9.3 $10^{42}$ protons per second. Dividing this number by the area of that sphere, we derive a proton flux of 1.7 $10^{11}$ $m^{-2}s^{-1}$ due to blobs during the period of study. If we assume that the proton flux of the solar wind at 1 AU is 2.5 $10^{12}$ $m^{-2}s^{-1}$ (Wang 2010, Sanchez-Diaz et al. 2016), and that 40 to 50% of the solar wind is SSW during solar maximum (Tokumaru et al. 2010), we estimate that the proton flux released into the solar wind in the form of blobs would represent around 15% of the bulk of the whole SSW at solar maximum.

Based on it-situ measurements, Kasper et al. (2007) and McGregor (2011) suggest that there are two sources of SSW. A low He abundance solar wind is detected near the current sheet and a SSW, with similar composition to the fast solar wind, is detected at higher heliographic latitudes, probably originating at the boundaries of large coronal holes (Wang et al. 2009). The low-He component of the SSW constitutes around 10% of the SSW at the maximum of solar cycle 23, and around 50% of the SSW at its minimum (Kasper et al. 2007; McGregor 2011). The latitudinal distribution of the two components of the SSW during solar cycle 24 is not as apparent as during solar cycle 23, due to a generally higher tilt of the HCS and the fact that these components are more difficult to separate statistically during cycle 24 (Michael Stevens & Justin C. Kasper, private communication). It is therefore difficult to estimate, from in-situ measurements, the percentage of SSW whose source is near the neutral line during our period of study. However, drawing parallels with results of the analysis of in-situ measurements from solar cycle 23, we conclude that our results could be consistent with the hypothesis that blobs constitute the low He abundance source of the SSW, associated with the current sheet.

If the hypothesis that blobs constitute the low He abundance source of the SSW is true, then either blobs carry a higher mass flux during solar minima than during solar maxima or the mass flux released by the high-latitude source of SSW is significantly decreased at solar minimum. However, the flat nature of the heliospheric current sheet during solar minima prevents blobs from being observed at high heliographic latitudes during those times. This makes it very difficult to analyze the variability along the neutral line during solar minima without the aid of an out-of-ecliptic imager. This difficulty will be overcome, to some degree, in the next years thanks to the inclined orbit of Solar Orbiter. In addition, the Parker Solar Probe will provide additional in-situ data over a large range of heliospheric altitudes, facilitating the simultaneous in-situ sampling of blobs.

# 4. CONCLUSIONS

In this paper, we have characterized the time variability of the heliospheric current sheet as well as its spatial variability during Carrington rotation 2137, during the maximum of solar cycle 24. We found that the neutral line releases a profusion of blobs, with a typical periodicity of 19.5 hours, simultaneously at all latitudes along its full extent. The size of such blobs is around 12x5 $R_s$ in the plane containing the HCS when it reaches a heliocentric distance of 30 $R_s$. This size is much bigger than the apparent size of the signature of blobs in running-difference images (1x0.1 $R_s$; Sheeley et al. 1997). As noted in previous sections, such signatures in running-difference images track the edges of blobs rather than the blobs themselves. Background-subtracted images provide a better estimate than running-difference images for the actual size of blobs and for their spatial distribution in the heliosphere. For this reason, previous studies that used running-difference images (e. g. Sheeley et al. 1997, Wang et al. 1998, Song et al. 2009) may have overestimated the frequency of release of blobs and underestimated their size.

The spatial and temporal distribution of blobs can also be used to estimate the mass flux released into the SSW in the form of blobs. Assuming that the proton density of a blob is 10 $cm^{-3}$ at 1 AU, we find that the flux associated with blobs during this event is $1.7\ 10^{11}\ m^{-2}$. This represents 15% of the SSW. This would be sufficient to account for the low latitude low He component of the SSW released during the activity maximum of solar cycle 23 (Kasper et al. 2007, McGregor et al. 2011), although our results, of course, correspond to the solar maximum of solar cycle 24.

Blobs were first observed by Sheeley et al. (1997), in running-difference images, as an intermittent release of brightness enhancements that propagate radially outwards. Sheeley et al. (2009) demonstrated that such point sources of light were the projection of small-scale expanding loops. Also using running-difference images, Rouillard et al. (2010ab) associated the precise arrival time of blobs in-situ to the trailing edge of magnetic clouds. This suggested that the expanding bright loops observed by Sheeley et al. (2009) could be density accumulated near the trailing edge of flux ropes and/or in the reconnection X-line likely involved during the formation of the flux rope. This is supported by the existence of inward moving plasma in the solar corona (e. g. Sheeley et al. 2001, 2014), whose formation is systematically associated to the outward release of blobs (SD17). SD17 interpreted this as strong suggestion that magnetic reconnection takes place around 5-6 $R_s$ from Sun center in order for the blob to be released. Note that the observations presented in SD17 were performed for solar maximum conditions. It is, thus, possible that the height of this reconnection could vary with the solar cycle. This hypothesis is supported by the cyclic variation of the height and shape of the source surface that defines approximately the height at which open magnetic fields of opposite polarity meet to form the coronal neutral sheet (e. g. Riley et al. 2006). As stated in the discussion of SD17, the coronagraph observations do not preclude the formation of blobs also at lower heights. In the present paper, we quantify the characteristic spatiotemporal scales associated to the release of blobs and estimate the mass flux injected into the SSW by the release of blobs near solar maximum

The analysis in this paper has been performed for a highly tilted neutral line that arose during the solar maximum of cycle 24. We note that observations from the upcoming out-of-ecliptic Solar Orbiter mission will provide the opportunity to repeat this study for the case of east-west directed neutral lines, that will likely form during the next solar minimum. This will allow an evaluation of whether the spatial and temporal scales estimated here for the release of blobs depend on the particular geometry of the coronal magnetic field.

# Acknowledgements
We acknowledge use of the tools made available by the plasma physics data center (Centre de Données de la Physique des Plasmas; CDPP; http://cdpp.eu/), the Virtual Solar Observatory (VSO; http://sdac.virtualsolar.org), the Multi Experiment Data & Operation Center (MEDOC; https://idoc.ias.u-psud.fr), the French space agency (Centre National des Etudes Spatiales; CNES; https://cnes.fr/fr) and the space weather team in Toulouse (Solar-Terrestrial Observations and Modelling Service; STORMS; https://stormsweb.irap.omp.eu/). This includes the data mining tools AMDA (http://amda.cdpp.eu/), CLWEB (clweb.cesr.fr/) and the propagation tool (http://propagationtool.cdpp.eu). RFP, IP, ESD, JAD and EK acknowledge financial support from the HELCATS project under the FP7 EU contract number 606692. The STEREO SECCHI data are generated by a consortium comprising RAL (UK), NRL (USA), LMSAL (USA), GSFC (USA), MPS (Germany), CSL (Belgium), IOTA (France) and IAS (France). The Wind data were obtained from the Space Physics Data Facility. We also acknowledge the use of SDO AIA and HMI data

# Appendix 1: Coordinates

A heliospheric image is a projection of a portion of the heliosphere onto a 2D surface. The goal of the change of coordinates explained here is to aid in the recovery of the 3D structure of the heliosphere from heliospheric images. For this purpose, first, we introduce a 2D projected coordinate system centered on the observer and a 3D coordinate system centered on the Sun. Finally; we derive the equations for the change of coordinates between these two reference frames given certain assumptions. For a detailed description of these, and other widely-used coordinate systems in heliophysics, see Thomson (2006).

Only two coordinates are needed to define the position of a pixel in an image. In the Helio-projected radial (HPR) coordinate system, these two coordinates are the elongation, $\varepsilon$, and the position angle, PA. Let us define a point P in the heliosphere, whose coordinates will be described. The Line of Sight (LOS) that observes point P is defined as the line between the observer and the point P. The elongation, $\varepsilon$, of point P is the angular distance in the image between the point and Sun center, i. e., the angle between the LOS corresponding to point P and the line joining the Sun and point P. The position angle, PA, of point P is defined as the angle between the projection of that P in the image and the projection of the rotation axis of the Sun in the image (and is measured counterclockwise from solar equatorial north). These two coordinates are illustrated in Figure A1a, on a COR2-A image. The elongation is also illustrated in Figure A1b.

On the other hand, three coordinates are needed to define the position of a point in 3D space. For this purpose, we use the HEEQ coordinate system in this paper. The HEEQ coordinate system is illustrated in Figure A1b, together with other useful coordinates, such as elongation and the β angle. The radial coordinate, r, is defined as the distance between point P and the center of the Sun, the heliographic latitude, $\lambda$, is defined as the angle between the line from the point P to Sun-center and the solar equator. $\lambda$ is measured on the meridian containing point P, i. e., the plane containing point P and the solar rotation axis. The heliographic longitude, l, is the angle between the line joining the projection of point P on the solar equatorial plane (P'') to the Sun-center (line SP'' in Figure A1b) and the y axis (the line joining the Sun and the projection of Earth onto the equatorial plane).

It is also useful for our calculations to define the β angle, which is the angle between the observer-P line (line AP) and observer-Sun center line, AS (the observer in our case being STEREO-A). Like the elongation, β is measured on the plane containing the observer (A), point P and Sun-center (S). Note that this quantity is, in general, not the same as the difference in longitude between the observer and point P because the longitude and the β angle are, in general, not measured on the same plane (see Figure 12 of Rouillard et al. 2009).

Each pixel detects the integrated white light along its LOS, defined by its coordinates (e.g. $\varepsilon$, PA). This LOS contains an infinite number of points in 3D space, each one defined by a set of coordinates (e.g. r, $\lambda$, l). We make the assumption that all the white light observed in a pixel has been scattered by a blob. We further assume that blobs have their origin at the neutral line (Plotnikov et al. 2016) and move at constant velocity radially outwards. In the case of a highly tilted neutral line, this places the blobs on a 3D spiral centered on the Sun corotating with the neutral line. With these two assumptions, given a pair of values ($\varepsilon$, PA) and the time (t) when the image is taken, there is a unique conversion between 2D HPR coordinates ($\varepsilon$, PA) and 3D HEEQ coordinates (r, $\lambda$, l). The point observed along a LOS is the intersection between the LOS from the corresponding pixel and the Parker spiral; the latter defined by the period of rotation of the Sun and the speed of the SSW. As the spiral corotates with the neutral line, the 3D coordinates of the point in the heliosphere that is observed along a particular LOS varies with time.

In order to construct a LAT-map, we select a fixed height above the center of the Sun, r, and evaluate which pixels, defined here in terms of (PA, $\varepsilon$), correspond to each set of 3D locations (r, t, $\lambda$). In reality, as we bin the data to produce the LAT-map, we pick a small range of 3D locations and determine the pixels, the LOSs from that fall within that range. For the LAT-map presented in Figure 4a, we select r as 30$R_s$, select images (each having a constant time) between 2013 May 30 to 2013 June 3, and define the HEEQ latitude ($\lambda$) as ranging from -40° to 40°. The HEEQ longitude of the point of release of each blob, which we assume to be maintained along its full trajectory, can be calculated using the angular velocity of the solar rotation, $\omega$, the position of the neutral line at time $t_0$, when it was aligned with Earth, and the speed of the blob, v:

$$l = \omega\left(t - t_0 - \frac{r}{v}\right) \quad (A1)$$

This equation associates, with each image, a relation l=l(r). Since r is fixed for a given LAT-map, the equation provides, for each image, the HEEQ longitude corresponding to this height. Equation 10 of Rouillard et al. (2009) can be used to convert each latitude/longitude ($\lambda$, l) pair into $\beta$:

$$\sin^2(\beta) = \cos^2(\lambda)\sin^2(l) + \left[-\cos(\lambda)\cos(l)\sin(A_0) + \sin(\lambda)\cos(A_0)\right]^2 \quad (A2)$$

where $A_0$ is the HEEQ latitude of the observer (STEREO-A). This value of $\beta$ is used in Equation 11 of Rouillard et al. (2009) to retrieve the elongation angle:

$$\tan(\varepsilon) = \frac{r\sin(\beta)}{r_A - r\cos(\beta)} \tag{A3}$$

and in Equation 1 of Savani et al. (2012) to compute PA:

$$\cos(PA) = \frac{\cos(\beta)\sin(A_0) - \sin(\lambda)}{\sin(\beta)\cos(A_0)} \tag{A4}$$

Hence, this algorithm enables transformation between the 3D coordinates of a blob (r, t, $\lambda$) and its 2D projection (PA, $\varepsilon$). The resulting values of (PA, $\varepsilon$) for each time determine the pixels that are included into the LAT-map of a given r.

Based on these equations, Figure A2 shows the elongation (left panel) and PA (right panel) corresponding to each pair of values (here time and latitude) in the (x, y) axis for a fixed height above the Sun of 30 $R_s$. Note that the elongation varies by a few degrees over time at a given latitude. Contrary to coronagraphs, which only observe features near the plane of sky, heliospheric imagers, due to their higher sensitivity, observe over a large range of longitudes. This provides the opportunity to observe the blobs entrained in the CIR for nearly a week. Therefore, it is not justified to assume that the blobs lie in the Thomson sphere and we need to use the transformation described above (e. g. Howard & DeForest 2012). The PA corresponding to a fixed latitude also varies significantly with time, and it differs by a number of degrees from the heliocentric ecliptic colatitude due to the fact that STEREO is, for most of the time, not in the solar equatorial plane.

# FIGURES

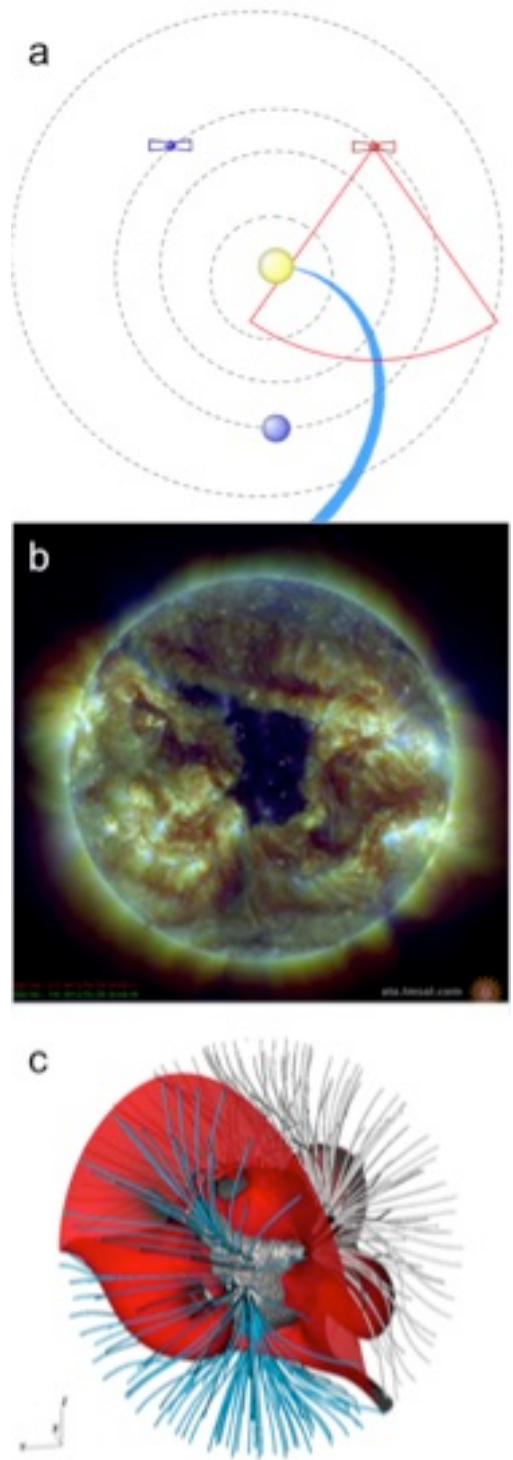

**Figure 1.** (a) Polar map of the ecliptic plane showing the position of Earth (blue dot; not to scale), Sun (yellow dot; not to scale), STEREO-A (red spacecraft symbol; not to scale) and STEREO-B (blue symbol; not to scale), the FOV of STEREO HI-A (red triangle), and the CIR (blue spiral) on 2013 June 3 with an arbitrary thickness. Plot generated with the propagation tool (http://propagationtool.cdpp.eu). (b) AIA image of the solar corona (combined 193 and 211 Å) on 2013 May 29. (c) PFSS reconstruction of the coronal magnetic field for solar rotation 2137 with the neutral sheet (red sheet). Each line color

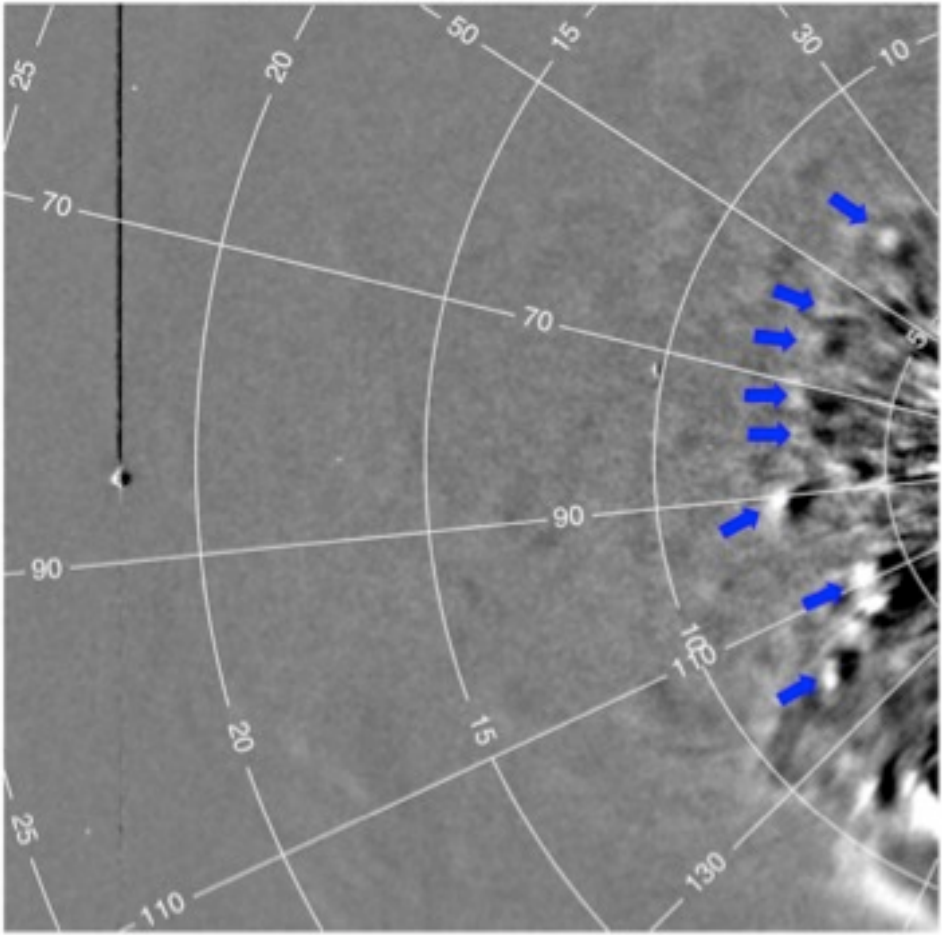

**Figure 2.** Smoothed running-difference image of STEREO HI1-A on 2013 May 31 18:49 UT. Image generated by RAL Space (http://www.stereo.rl.ac.uk/cgi-bin/movies.pl). Some of the clearest blobs are indicated with blue arrows. The white straight lines are drawn at constant PA. The number on each line is the PA in degrees.. The white arcs are drawn at constant elongation. Such elongation is indicated with a white number

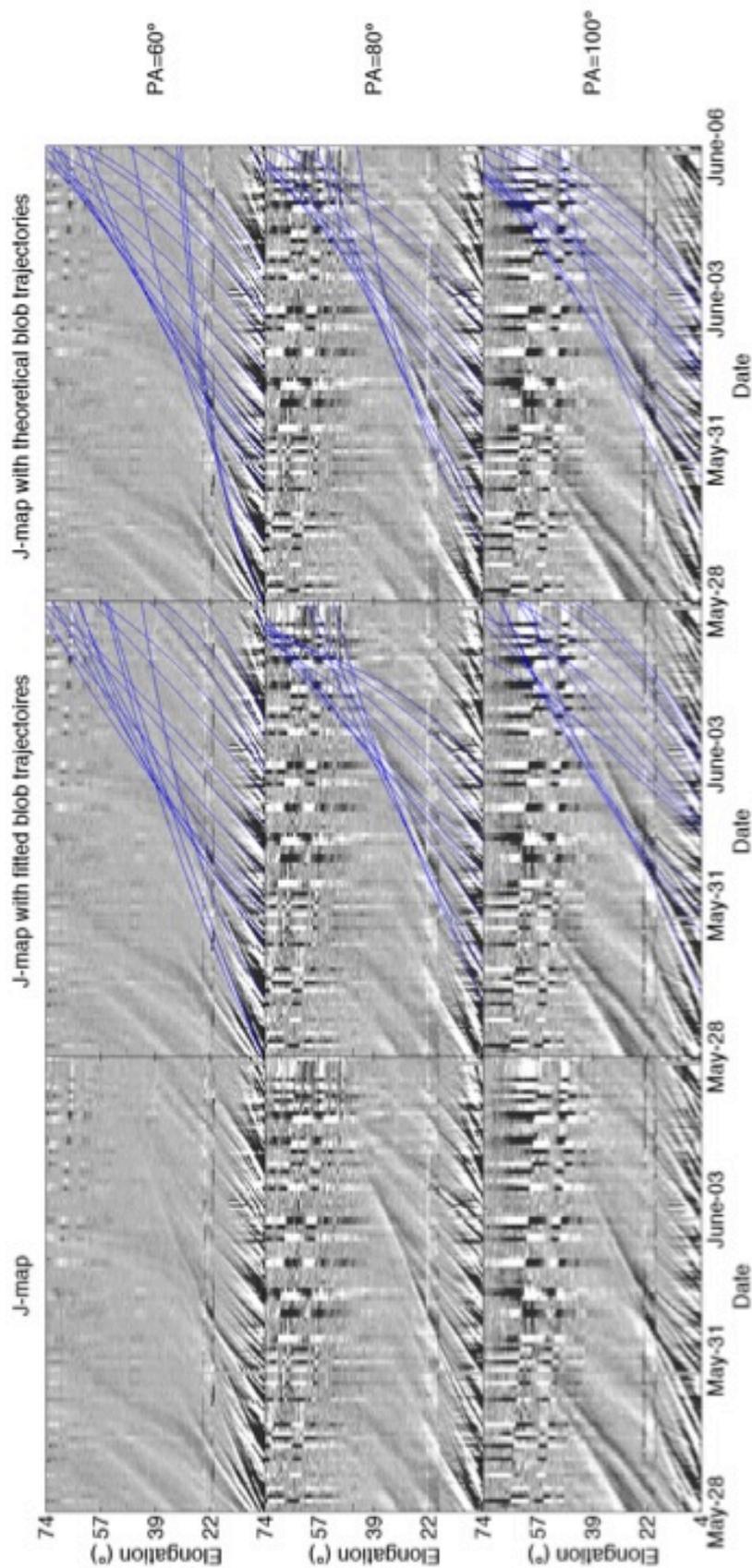

**Figure 3.** J-maps constructed from HI-1 and 2 running-difference data from STEREO-A at 60° (top row), 80° (central row), and 100° (bottom row) PA during the passage of the CIR, from 2013 May 28 to June 06. The blue lines over plotted in the middle column are the trajectories fitted to the manually-extracted time-elongation profile of each blob. The blue lines over plotted in the right column are the theoretical trajectories of blobs, derived assuming that they move radially at constant speed after having been created at the neutral line.

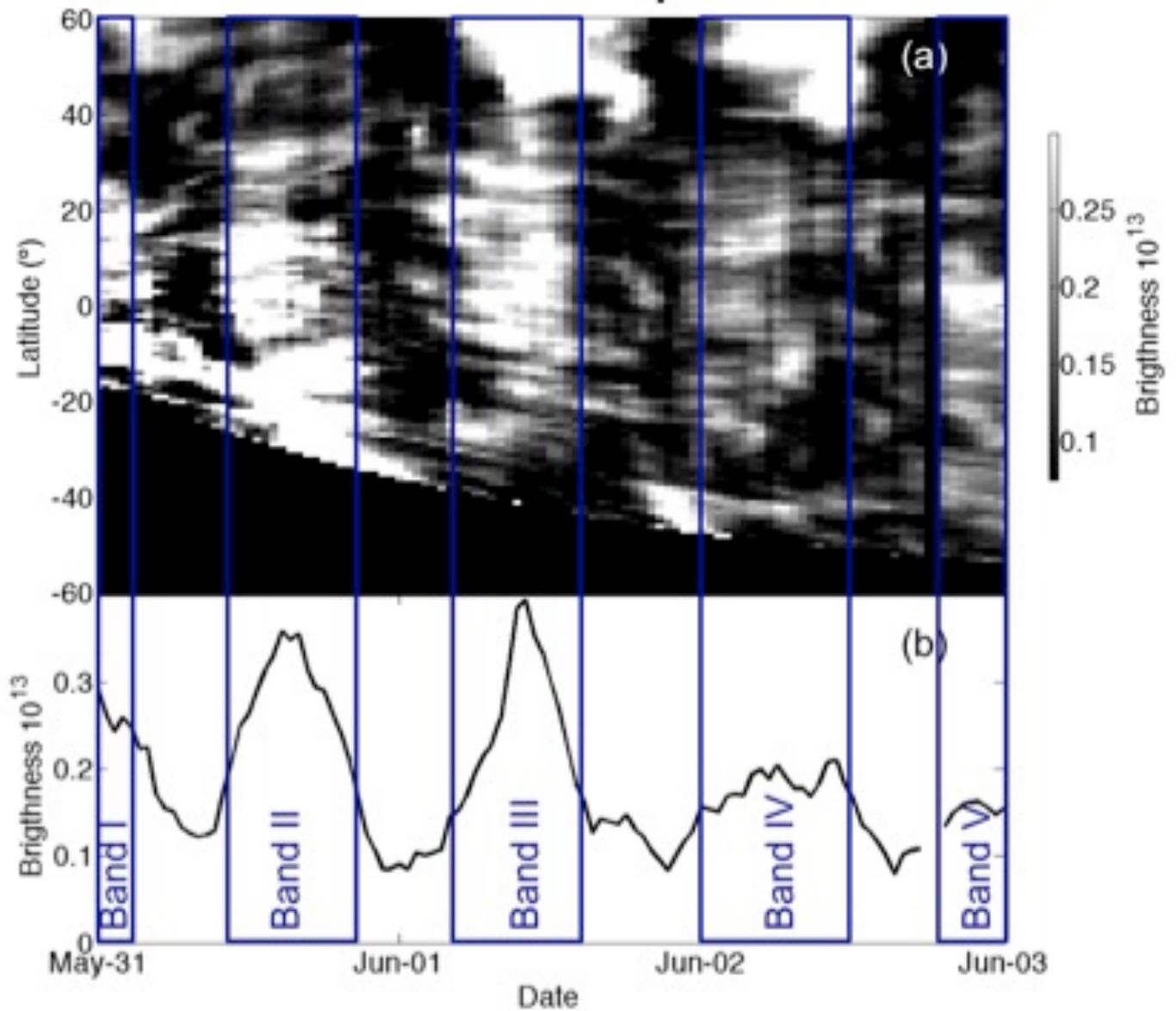

**Figure 4:** (a) LAT-map of STEREO HI-1A data at a fixed height of 30 $R_s$ above the Sun for the period from 2013 May 31 to June 03 constructed with 1-day background-subtracted images. (b) Time series of latitude-averaged brightness, averaged over all latitudes in the LAT-map at each point in time. The blue rectangles highlight five bands of increased brightness that extends over all observable latitudes.

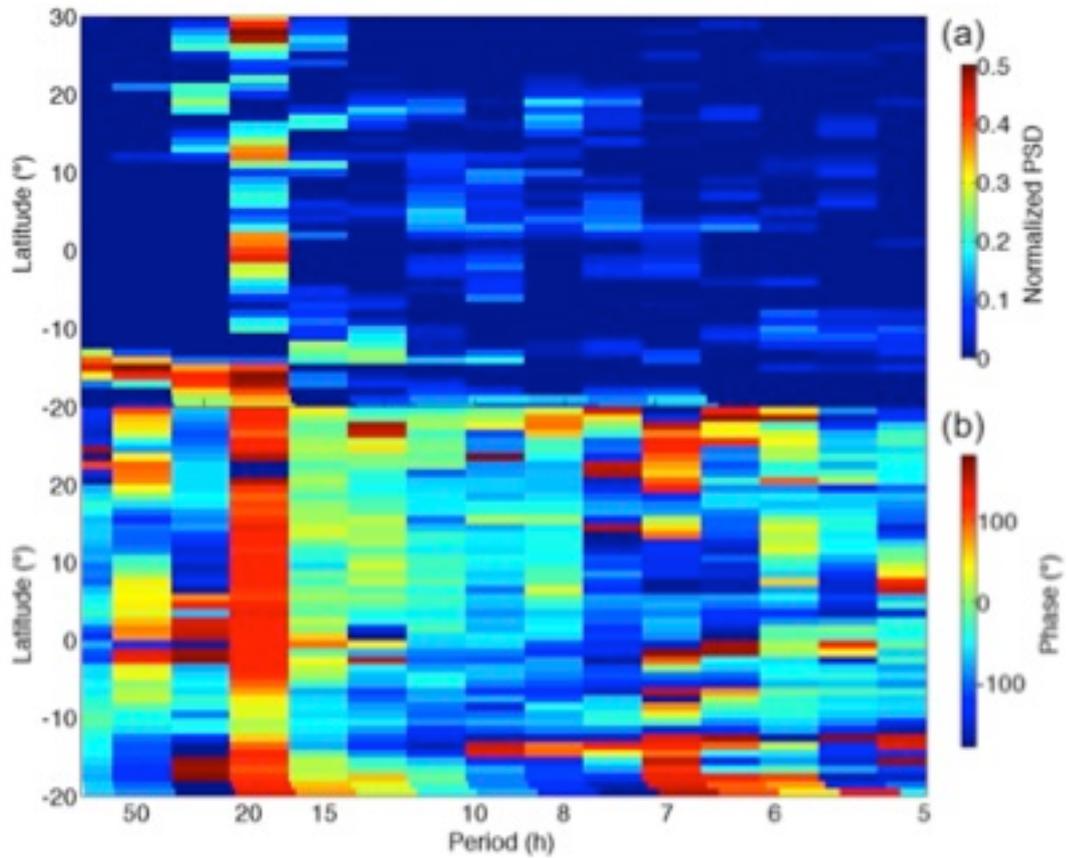

**Figure 5:** Magnitude (a) and phase (b) of the Fourier transform of the brightness-time series at each latitude in the LAT-map of STEREO-A HI-1 data at 30 $R_s$, presented in Figure 4b. A red noise with spectrum $S=S_0 \cdot f^{\alpha}$ has been subtracted from panel a.

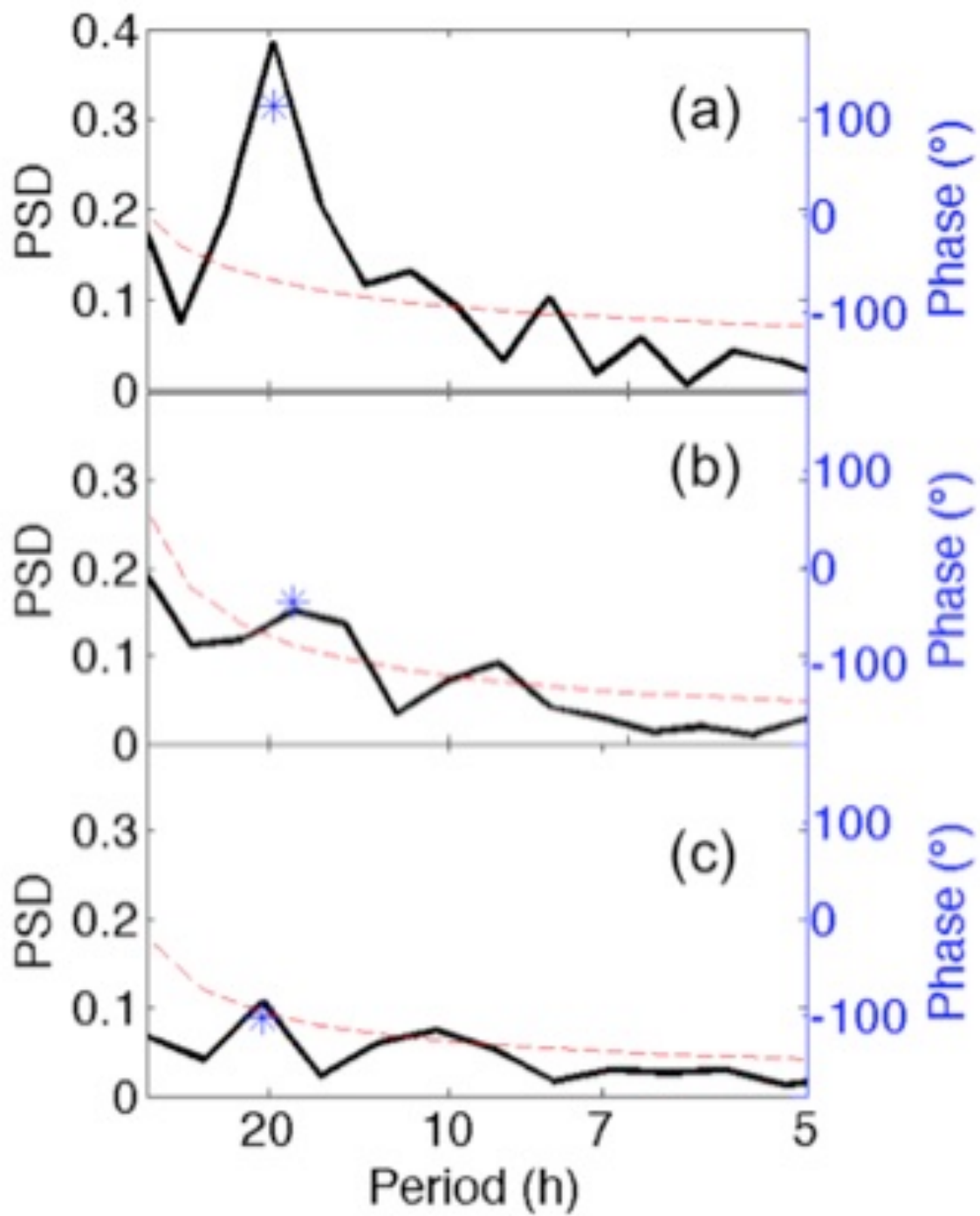

**Figure 6.** Frequency spectra (black solid lines) of the latitude-averaged brightness time series at 30 $R_s$ (a), 40 $R_s$ (b) and 50 $R_s$ (c), derived from the LAT-maps generated at each altitude. The phase of the Fourier transform corresponding to the spectral peak at 19.5-hours periodicity is shown as a blue asterisk. The red dashed line shows the 99% confidence level above a red noise, with spectrum $S=S_0 \cdot f^\alpha$, according to a chi-square test.

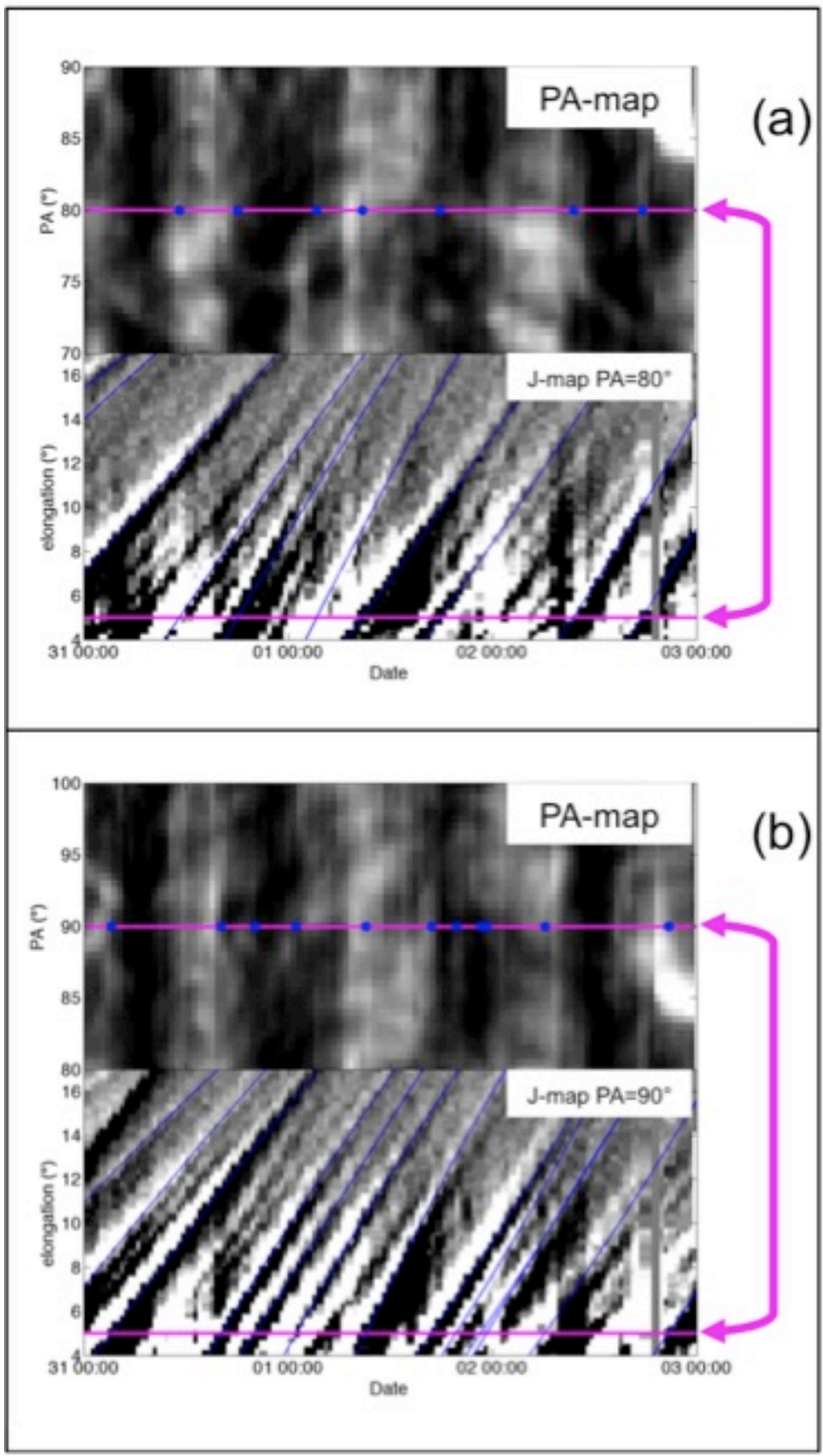

**Figure 7:** Each set of panels consists of a pair of maps. The top map of each pair is a PA-map at 5° elongation, centered at 80° PA (panel a) and 90° PA (panel b) with a 20° PA width, constructed from background-subtracted STEREO-A HI-1 data. The bottom map of each pair is a J-map from 4° to 17° elongation constructed at the central PA of the corresponding PA-map but with running difference data. The blue lines are the trajectories of brightness variations observed in running-difference images calculated from their manually derived time-elongation profile. The blue points in the PA-map show the time of passage of each of these brightness enhancements through the elongation of 5°. Magenta

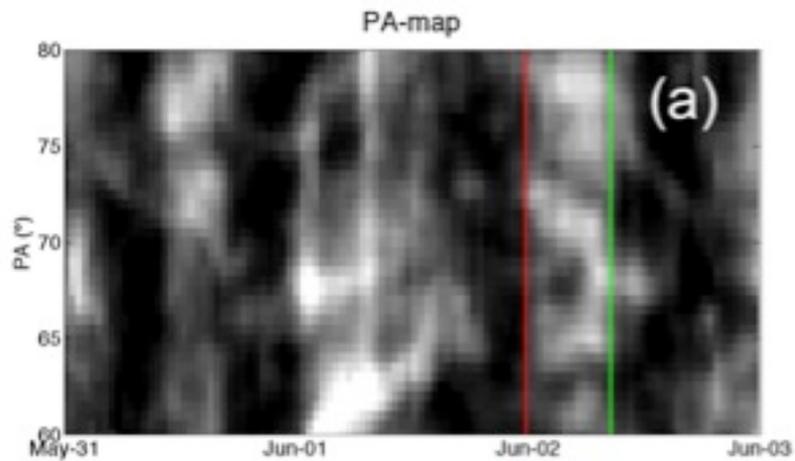

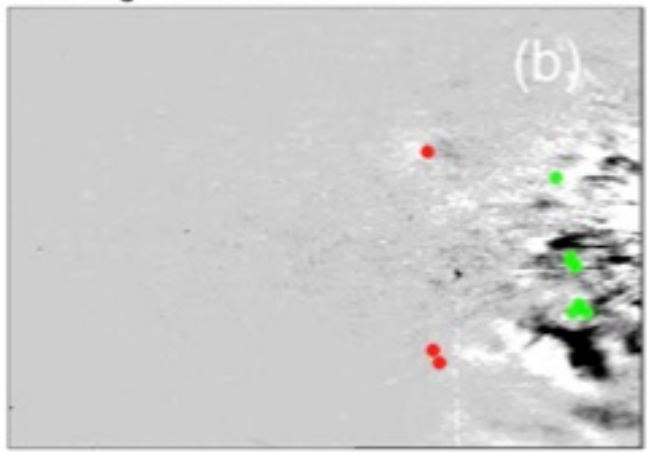

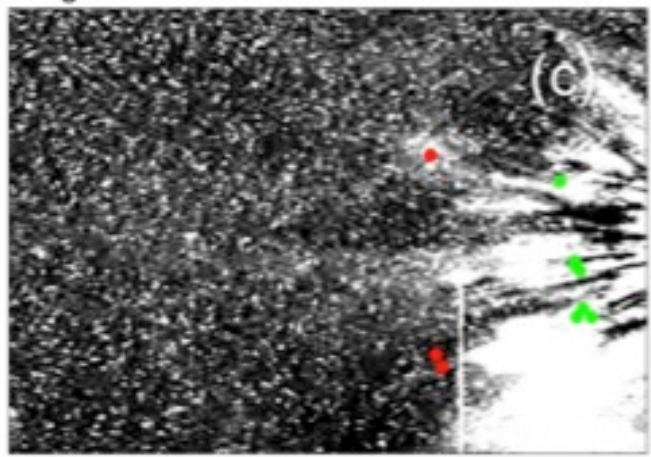

**Figure 8.** (a): PA-map centered at PA=70°. The red and green vertical lines mark the passage through 5° elongation (the elongation of this PA-map) of two outward-moving density features observed in a J-map. Portions of running difference (b) and background subtracted (c) STEREO-A HI-1 images on 2013 June 02 10:49 UT. The position of the density features corresponding to the red (green) line in the PA-map are marked with red (green) dots in each image.

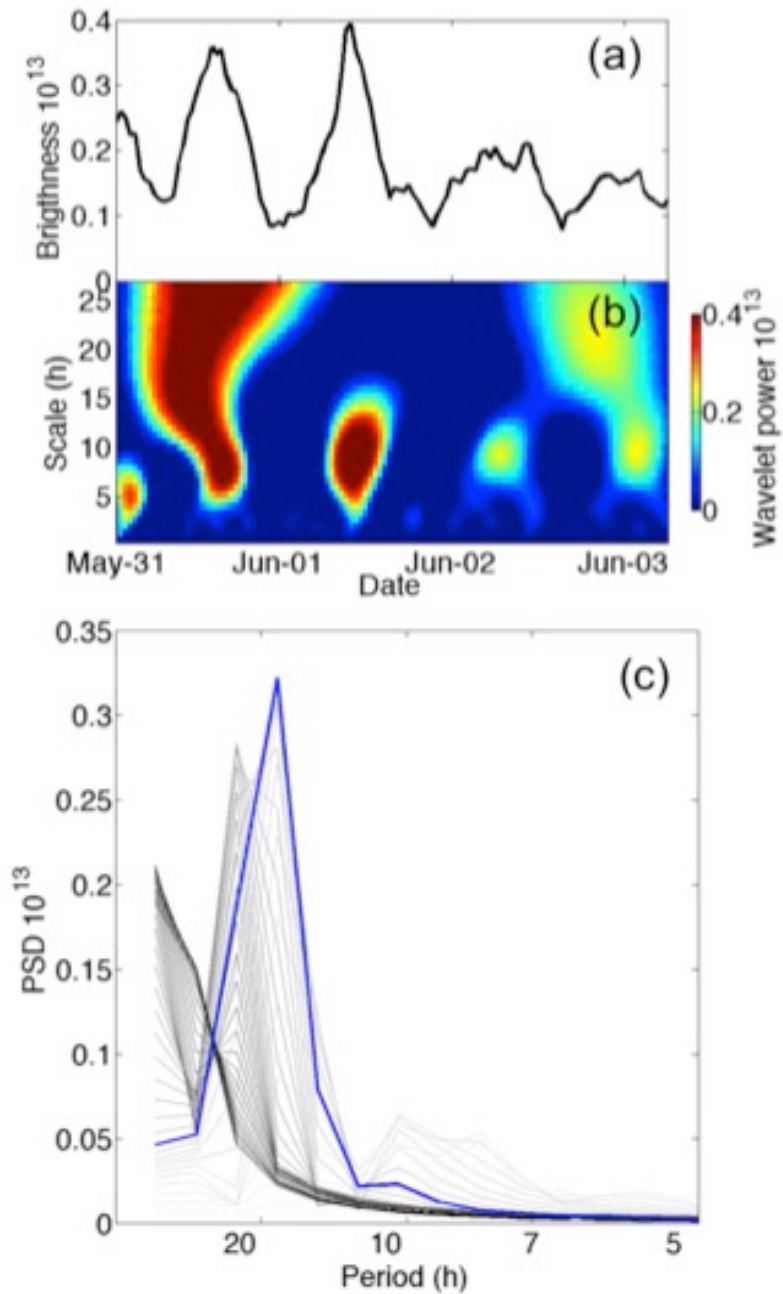

**Figure 9.** Time series of latitude-averaged brightness at 30 $R_s$ (repeated from Figure 4b). (b) Wavelet transform of the time series in panel a. (c): Frequency spectrum (Power Spetral Density; PSD) of the time series resulting from each wavelet decomposition, at each scale on panel b. The blue thick solid line is the spectrum for the wavelet decomposition at the scale 9 hours. The color of each line represents the wavelet scale in a gray scale from 40 minutes (white) to 26 hours (black).

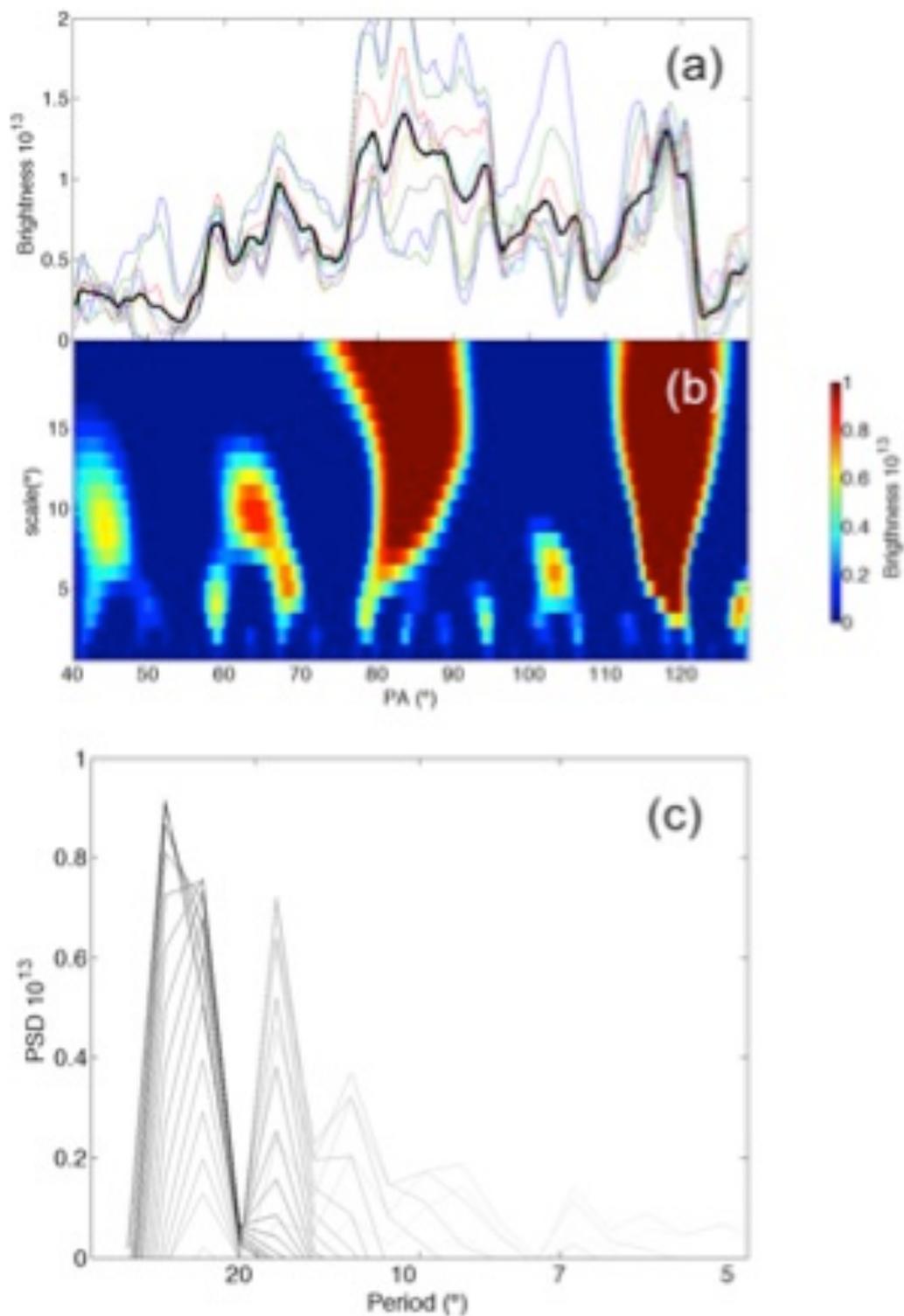

**Figure 10.** (a) PA-brightness profiles at 30 $R_s$ for the time interval 2013 May 30 18:00 UT to 31 00:00 UT. The colored lines show the brightness-PA profile at each given time and the black thick line the average of all of them. (b) Wavelet transform of the averaged brightness-PA profile shown in panel a. (c) Frequency spectra (Power Spetral Density; PSD) of each scale of the wavelet transform of the PA series. The color of each line correspond to the wavelet scale in a gray scale from 1° (white) to 20° (black).

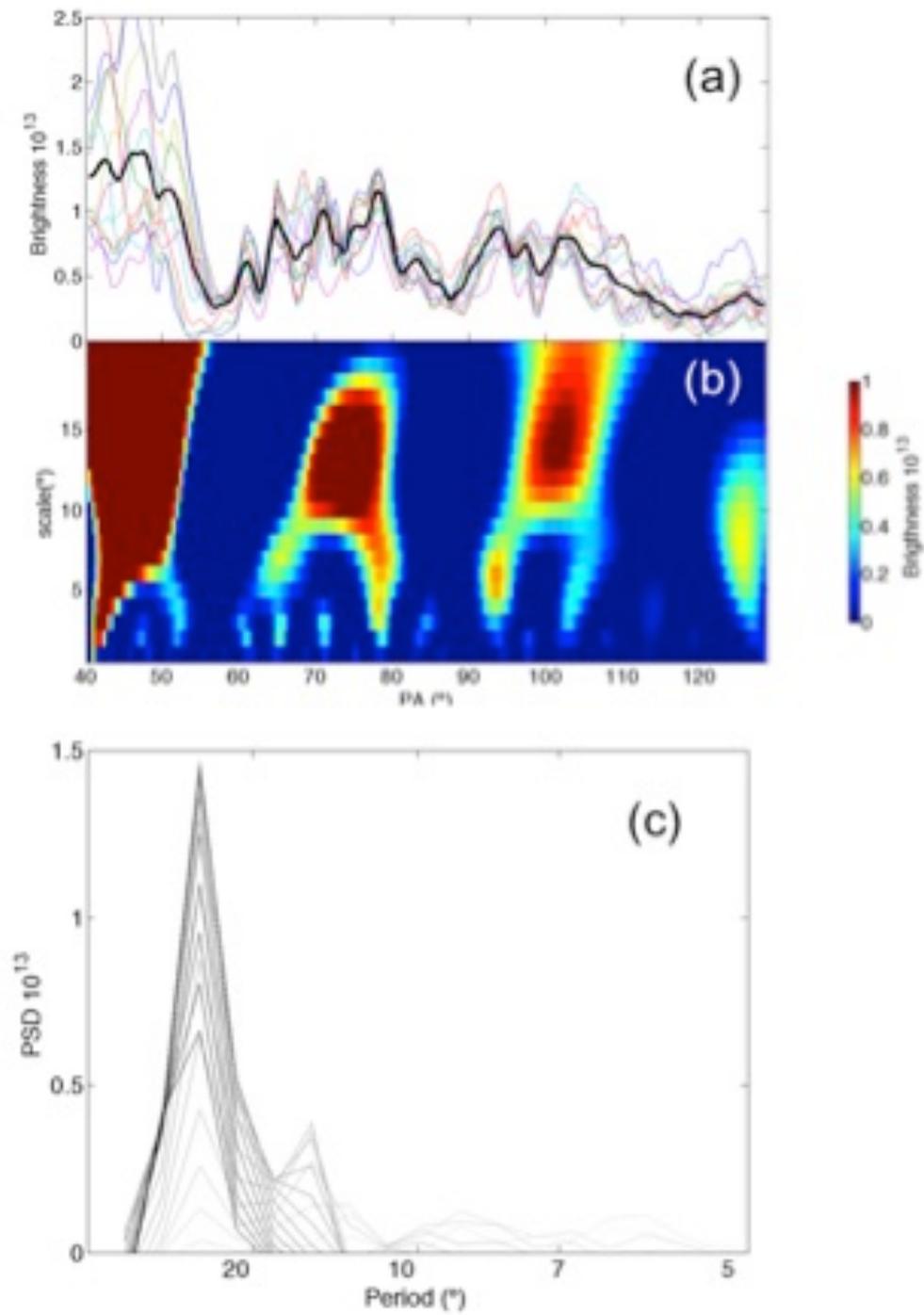

**Figure 11.** Same as Figure 10, but for the time interval extending from 2013 June 02 00:00 UT to 08:00 UT.

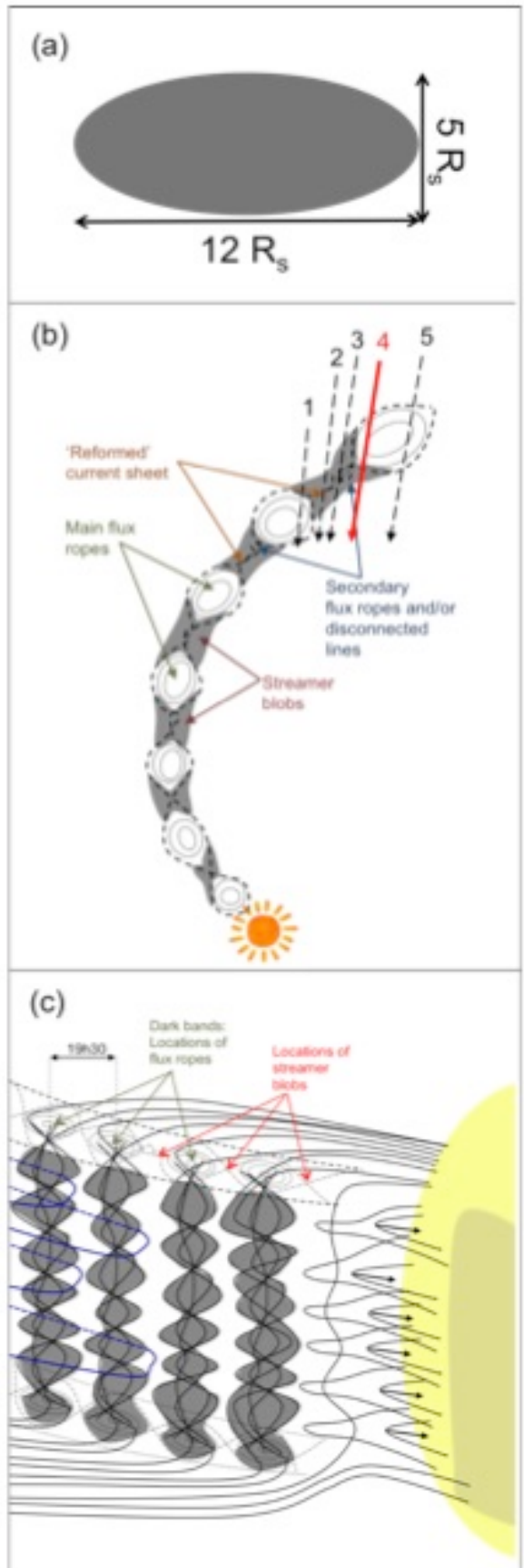

**Figure 12:** (a) Shape and size of the latitudinal cut of the density enhancement (blob) associated with a small transient. (b) View from above of the ecliptic plane with blobs, in dark gray, released by an exactly corotating source. The Sun, center bottom, is not to scale but the size of the blobs and the distances are to scale. The black arrows sketch different possible types of trajectories for a spacecraft relative to the position of blobs. (c) Latitudinal cut of the heliosphere with its magnetic structure and with blobs in dark grey. The Sun, on the right side, is not to scale.

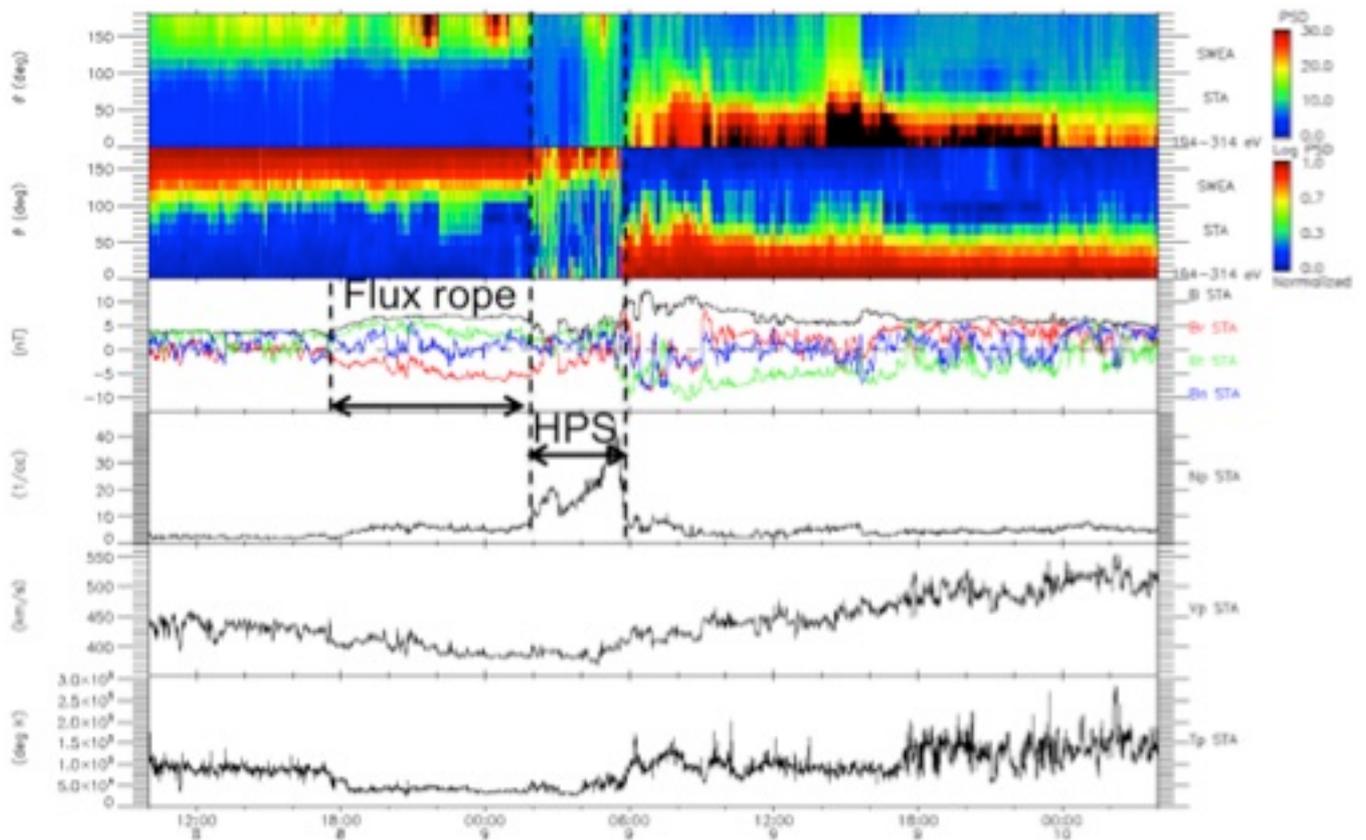

**Figure 13:** In-situ measurements of STEREO-A from 2013 July 08 12:00 UT to 10 00:00 UT. From top to bottom: suprathermal electron pitch angle distribution, normalized suprathermal electron pitch angle distribution, interplanetary magnetic field in RTN coordinates, proton density, proton bulk speed, proton temperature. The black dashed lines and the arrows indicate the passage of a flux rope and a high-density region, labeled HPS.

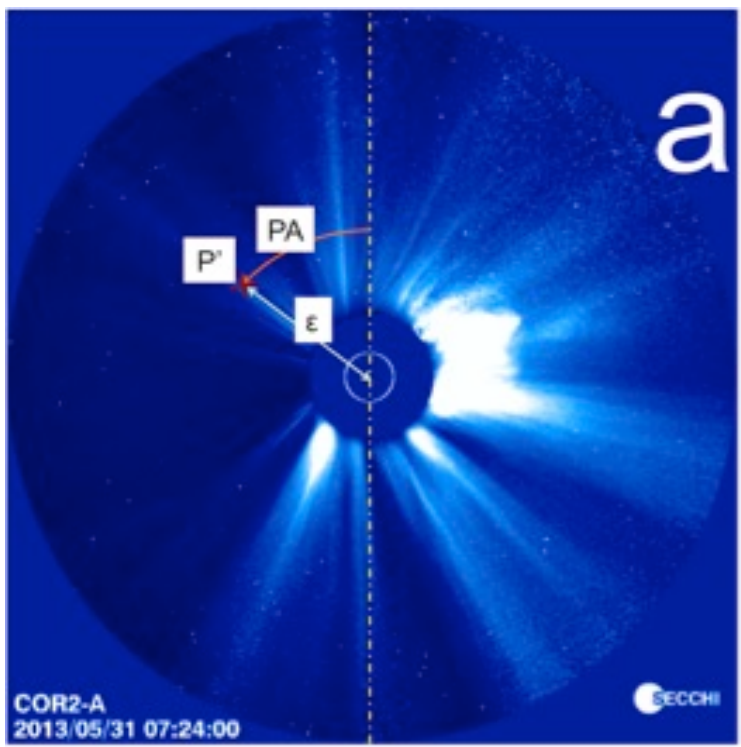

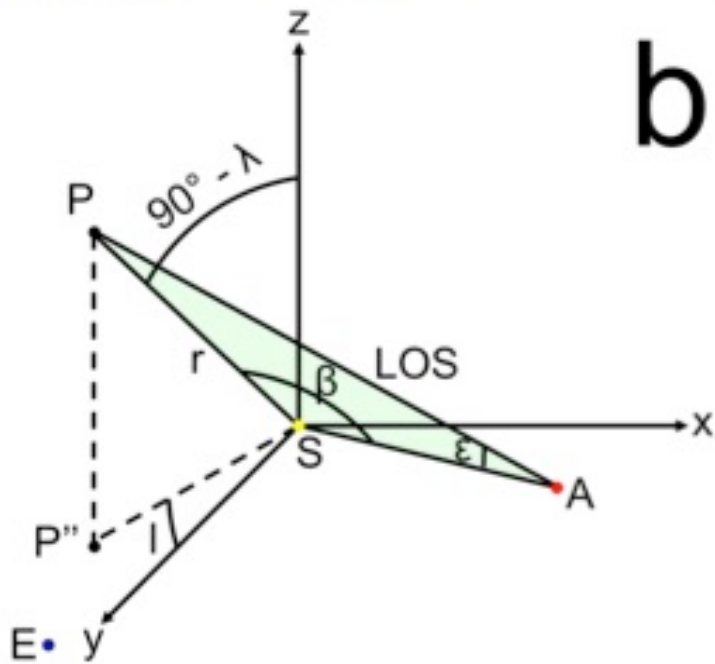

**Figure A1:** sketch of the helio-projected radial coordinate system and the HEEQ system. (a): COR2-A image of the solar corona on 2013 May 31 at 07:24 UT. The vertical dashed yellow line is the projection of the solar rotation axis of the Sun in the image. The red X indicates the projection in the image (P') of the point P, whose coordinates are defined in the Figure. PA stands for position angle, which is measured counterclockwise from the projection of the rotation axis of the Sun to the line between P' and the projection of the center of the Sun in the image. ε stands for the angular distance from the Sun center as seen from STEREO-A to the projection of the point P in the image. (b): Coordinates placed in a reference system centered on the Sun. In the HEEQ coordinate system, the origin is set at the Sun center (S), the z axis is defined as the direction of the solar rotation axis and the y axis points towards the projection of the of the Earth in the solar equatorial plane. The red point (A) denotes the position of the observer, the point P is the point whose coordinates we are defining and P'' is its projection into the solar equatorial plane. The line between A and P, labeled LOS, is the line of sight that passes through the point P. The distance between S and P is the HEEQ radial coordinate, r. The angle between this line and the solar equator is the HEEQ latitude, λ. The angle between the SP line and the solar rotation axis is, thus, 90°-λ. The angle between line SP'' and the y axis is the HEEQ longitude, l. The angle ε, also defined in panel a, is the angle between the Sun-observer line and the observer-point line. We also define the β angle in this plot, which is the angle between the line SA and the line SP. Note that these two angles are measured in the plane containing the point P, the observer and the Sun, shaded in green. This plane is, in general different from the solar equator and from any solar meridian.

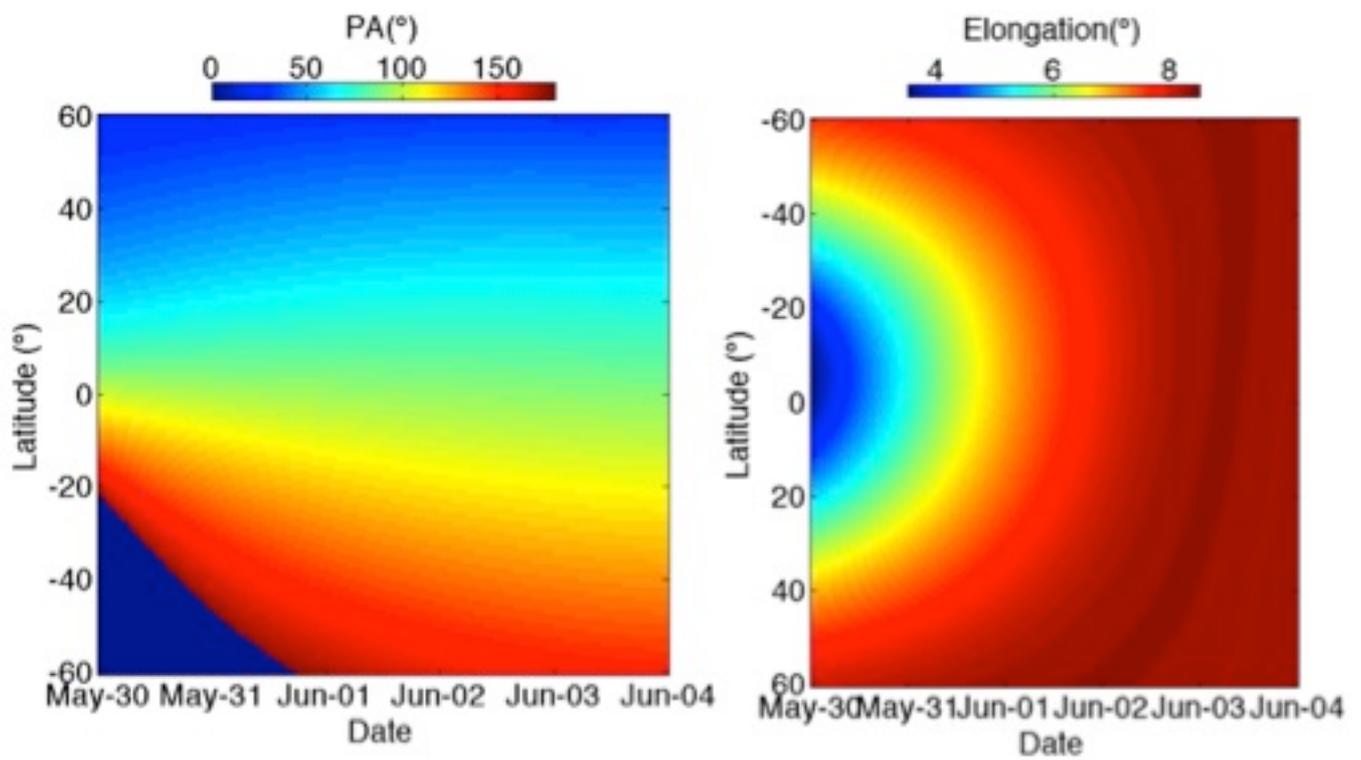

**Figure A2:** Change of coordinates from latitude, time of the image to PA (left) and to elongation (right)

| Band I (May 30 18:00 UT to 31 00:00 UT) | | |
|---|---|---|
| Scale | 8° | 15°-20° |
| Periodicity | 18° | 29°-44° |
| Band II (May 31 12:00 UT to 16:00 UT) | | |
| Scale | 11° | 17° |
| Periodicity | 22° | 29°-44° |
| Band III (June 01 06:00 UT to 10:00 UT) | | |
| Scale | 10° | - |
| Periodicity | 22° | |
| Band IV (June 02 00:00 UT to 08:00 UT) | | |
| Scale | 7°-8° | 15°-20° |
| Periodicity | 15° | 29°-44° |
| Band V (June 02 20:00 UT to 03 00:00 UT) | | |
| Scale | 8° | 16° |
| Periodicity | 15°-18° | 29° |

**Table 1.** Dominant periodicities in the spectral analysis of the wavelet transform of the averaged PA-brightness profile averaged for all times contained within each bright vertical band of the PA-map.